\documentclass{emulateapj}

\usepackage{natbib}
\usepackage{amsmath}

\newcommand{\ubvri}{\protect\hbox{$U\!BV\!RI$} }

\newcommand{\bvz}{\protect\hbox{$BV\!z$} }
\newcommand{\about}{$\sim\!\!$~}

\newcommand{\be}{\begin{displaymath}}
\newcommand{\ee}{\end{displaymath}}

\def\lsim{\hbox{\rlap{\raise 0.425ex\hbox{$<$}}\lower 0.65ex\hbox{$\sim$}}}
\def\gsim{\hbox{\rlap{\raise 0.425ex\hbox{$>$}}\lower 0.65ex\hbox{$\sim$}}}

\def\arcsec{\hbox{$^{\prime\prime}$}}

\newcommand{\kms}{km~s$^{-1}$ }

\newcommand{\mean}[1]{\left \langle #1 \right \rangle}
\newcommand{\iue}{\protect{\it IUE}}
\newcommand{\hst}{\protect{\it HST}}
\newcommand{\stis}{\protect{\it STIS}}
\newcommand{\uvot}{\protect{\it UVOT}}
\newcommand{\jwst}{\protect{\it JWST}}
\newcommand{\hstiue}{\protect{\it HST/IUE}}

\shorttitle{Constraining Evolution in SNe~Ia}
\shortauthors{Foley et al.}

\begin{document}

\title{Constraining Cosmic Evolution of Type~Ia Supernovae}

\author{
{Ryan~J.~Foley}\altaffilmark{1}, %
{Alexei~V.~Filippenko}\altaffilmark{1}, %
{C.~Aguilera}\altaffilmark{2}, %
{A.~C.~Becker}\altaffilmark{3}, %
{S.~Blondin}\altaffilmark{4}, %
{P.~Challis}\altaffilmark{4}, %
{A.~Clocchiatti}\altaffilmark{5}, %
{R.~Covarrubias}\altaffilmark{3}, 
{T.~M.~Davis}\altaffilmark{6}, %
{P.~M.~Garnavich}\altaffilmark{7}, %
{S.~Jha}\altaffilmark{1,8}, %
{R.~P.~Kirshner}\altaffilmark{4}, %
{K.~Krisciunas}\altaffilmark{9}, %
{B.~Leibundgut}\altaffilmark{10}, %
{W.~Li}\altaffilmark{1}, %
{T.~Matheson}\altaffilmark{11}, %
{A.~Miceli}\altaffilmark{3}, 
{G.~Miknaitis}\altaffilmark{12}, %
{G.~Pignata}\altaffilmark{13}, %
{A.~Rest}\altaffilmark{2,14}, %
{A.~G.~Riess}\altaffilmark{15,16}, %
{B.~P.~Schmidt}\altaffilmark{17}, %
{R.~C.~Smith}\altaffilmark{2}, 
{J.~Sollerman}\altaffilmark{6,18}, %
{J.~Spyromilio}\altaffilmark{10}, 
{C.~W.~Stubbs}\altaffilmark{4,14}, %
{J.~L.~Tonry}\altaffilmark{19}, 
{N.~B.~Suntzeff}\altaffilmark{2,9}, 
{W.~M.~Wood-Vasey}\altaffilmark{4}, and %
{A.~Zenteno}\altaffilmark{20} %
}
\email{rfoley@astro.berkeley.edu}

\altaffiltext{1}{Department of Astronomy, 601 Campbell Hall, University of California, Berkeley, CA 94720-3411}
\altaffiltext{2}{Cerro Tololo Inter-American Observatory, Casilla 603, La Serena, Chile}
\altaffiltext{3}{Department of Astronomy, University of Washington, Box 351580, Seattle, WA 98195-1580}
\altaffiltext{4}{Harvard-Smithsonian Center for Astrophysics, 60 Garden Street, Cambridge, MA 02138}
\altaffiltext{5}{Pontificia Universidad Cat\'olica de Chile, Departamento de Astronom\'ia y Astrof\'isica, Casilla 306, Santiago 22, Chile}
\altaffiltext{6}{Dark Cosmology Centre, Niels Bohr Institute, University of Copenhagen, Juliane Maries Vej 30, DK-2100 Copenhagen \O, Denmark}
\altaffiltext{7}{Department of Physics, University of Notre Dame, 225 Nieuwland Science Hall, Notre Dame, IN 46556-5670}
\altaffiltext{8}{Kavli Institute for Particle Astrophysics and Cosmology, Stanford Linear Accelerator Center, 2575 Sand Hill Road, MS 29, Menlo Park, CA 94025}
\altaffiltext{9}{Department of Physics, Texas A\&M University, College Station, TX 77843-4242}
\altaffiltext{10}{European Southern Observatory, Karl-Schwarzschild-Strasse 2, D-85748 Garching, Germany}
\altaffiltext{11}{National Optical Astronomy Observatory, 950 North Cherry Avenue, Tucson, AZ 85719-4933}
\altaffiltext{12}{Fermilab, P.O. Box 500, Batavia, IL 60510-0500}
\altaffiltext{13}{Departamento de Astronomia, Universidad de Chile, Casilla 36-D, Santiago, Chile}
\altaffiltext{14}{Department of Physics, Harvard University, 17 Oxford Street, Cambridge, MA 02138}
\altaffiltext{15}{Space Telescope Science Institute, 3700 San Martin Drive, Baltimore, MD 21218}
\altaffiltext{16}{Johns Hopkins University, 3400 North Charles Street, Baltimore, MD 21218}
\altaffiltext{17}{The Research School of Astronomy and Astrophysics, The Australian National University, Mount Stromlo and Siding Spring Observatories, via Cotter Road, Weston Creek, PO 2611, Australia}
\altaffiltext{18}{Department of Astronomy, Stockholm University, AlbaNova, 10691 Stockholm, Sweden}
\altaffiltext{19}{Institute for Astronomy, University of Hawaii, 2680 Woodlawn Drive, Honolulu, HI 96822}
\altaffiltext{20}{Univ. of Illinois, Dept. of Astronomy, 1002 W. Green St., Urbana, IL 61801}

\begin{abstract}
We present the first large-scale effort of creating composite spectra
of high-redshift type Ia supernovae (SNe~Ia) and comparing them to
low-redshift counterparts.  Through the ESSENCE
project, we have obtained 107 spectra of 88 high-redshift SNe~Ia with
excellent light-curve information.  In addition, we have obtained 397
spectra of low-redshift SNe through a multiple-decade effort at Lick
and Keck Observatories, and we have used 45 ultraviolet spectra obtained 
by \hstiue.  The low-redshift spectra act as a control sample
when comparing to the ESSENCE spectra.  In all instances, the ESSENCE
and Lick composite spectra appear very similar.  The addition of
galaxy light to the Lick composite spectra allows a nearly perfect
match of the overall spectral-energy distribution with the ESSENCE
composite spectra, indicating that the high-redshift SNe are more
contaminated with host-galaxy light than their low-redshift
counterparts.  This is caused by observing objects at all redshifts
with similar slit widths, which corresponds to different projected
distances.  After correcting for the galaxy-light contamination,
subtle differences in the spectra remain.  We have estimated the
systematic errors when using current spectral templates for
K-corrections to be \about 0.02 mag.  The variance in the composite
spectra give an estimate of the intrinsic variance in low-redshift
maximum-light SN spectra of \about 3\% in the optical and growing
toward the ultraviolet.  The difference between the maximum-light low and
high-redshift spectra constrain SN evolution between our samples to be
$< 10$\% in the rest-frame optical.
\end{abstract}


\section{Introduction}\label{s:intro}

Type Ia supernovae (SNe~Ia) are the most precise known distance
indicators at cosmological redshifts.  The meticulous measurement of
several hundred SNe~Ia at both low and high redshifts has shown that the
expansion of the Universe is currently accelerating
\citep{Riess98:lambda, Riess07, Perlmutter99, Astier06, Wood-Vasey07};
see \citet{Filippenko05} for a recent review.
The underlying assumption behind that work is that high-redshift
SNe~Ia have the same peak luminosity as low-redshift SNe~Ia (after
corrections based on light-curve shape; e.g.,
\citealt{Phillips93}).  The luminosity of a given SN and its
light-curve shape are determined by initial conditions of the white
dwarf progenitor star (e.g., mass at explosion, C/O abundance, and
metallicity), and the properties of the explosion (e.g.,
deflagration/detonation transition, the amount of unburned material,
and the density at the ignition point).  The progenitor properties are
set by the initial conditions at the formation of the progenitor
system, presumably having properties similar to the global galactic
properties at that time.  Since low-redshift SN~Ia progenitor systems
likely form, on average, in significantly different environments than
high-redshift SN~Ia progenitors, one may assume that some amount of
evolution is inevitable \citep[for a discussion of different causes
and effects of SN evolution, see][]{Leibundgut01}.

Theoretical studies of SN~Ia evolution have focused on the
composition, particularly metallicity, of the progenitor system as the
primary potential difference between the two samples.  There have been
two major studies with conflicting results.  For their study,
\citet{Hoflich98} changed the progenitor metallicity and modeled the
explosion, including a full nuclear-reaction network.  \citet{Lentz00}
changed the metallicity of the results of W7 models \citep{Nomoto84}
and input those parameters into their PHOENIX code
\citep{Hauschildt96} to produce synthetic spectra.  The main
difference between these methods is the definition of ``metallicity.''
\citet{Hoflich98} uses the term to mean the metallicity of the progenitor
star, while \citet{Lentz00} uses it to mean the metallicity of the
ejecta.

The differing definitions of metallicity yield different initial
conditions, which resulted in contradictory results from these
studies.  \citet{Hoflich98} suggest that with increasing metallicity,
the ultraviolet (UV) continuum of the SN increases, while \citet{Lentz00} 
suggest that it decreases. Ultimately, the differences are the result of
differing density structures \citep{Lentz00, Dominguez01}.  Although
the method of \citet{Lentz00} seems less physical than that of
\citet{Hoflich98} (simply scaling the metallicity of the ejecta by
solar abundances does not take into account, for example, that the
Fe-group elements are mainly produced in the SN explosion), they
provide model spectra for varying metallicities, which may elucidate
differences between low and high-redshift SN spectra.

Further predictions for lower metallicity include faster rise times
\citep{Hoflich98}, faster light-curve decline \citep{Hoflich98},
lower $^{54}$Fe production \citep{Hoflich98}, smaller blueshifting of
\ion{Si}{2} $\lambda 6355$ \citep{Lentz00}, decreasing $B-V$ color
\citep{Dominguez01, Podsiadlowski06}, and changing luminosity
\citep{Hoflich98, Dominguez01, Podsiadlowski06, Timmes03}.
\citet{Ropke04} suggest that the C/O ratio of the progenitor does not
significantly affect peak luminosity.

Observationally, a lack of evolution has been supported by
investigating various SN quantities such as rise time
\citep{Riess99:risetime}, line velocities \citep{Blondin06,
Garavini07}, multi-epoch temporal evolution \citep{Foley05}, line
strengths \citep{Garavini07}, and line-strength ratios \citep{Altavilla06}.
There have also been studies comparing the spectra of individual
high-redshift SNe~Ia to low-redshift SNe~Ia \citep{Riess98:lambda,
Coil00, Hook05, Matheson05, Balland07}, all of which have concluded
that there is no clear difference in spectral properties between the
two samples.

\citet{Bronder07} recently presented measurements of line strengths
that suggest a difference between low and high-redshift SNe~Ia in one
of three features measured.  They find that the difference is highly
dependent on the galaxy contamination at high redshift and might be
affected by their small low-redshift SN sample.  Consequently, they
note that the difference is interesting but not significant.

Despite the consistencies in spectral properties, \citet{Howell07}
note a slight shift in the mean photometric properties of SNe~Ia with
redshift.  They explain this evolution as a change in the ratio of
progenitors from the ``prompt'' and ``delayed'' channels
\citep{Scannapieco05}, corresponding to young and old progenitor
systems at the time of explosion, respectively.  In particular, the
light-curve shape parameter ``stretch'' \citep{Goldhaber01} increases
with redshift.  \hst\, observations of ESSENCE objects suggested that
the sample may have a large proportion of objects with slow-declining
(large stretch) light curves, but this is probably the result of a
selection bias\citep{Krisciunas05}.  Since stretch (and other
luminosity light-curve parameters) is correlated with spectral
properties, one might expect the spectra of high-redshift SNe~Ia, on
average, to differ from those of low-redshift SNe~Ia.

Since all galactic environments at redshift $0 < z < 1.5$ are also present 
in the local Universe, SN~Ia evolution does not necessarily mean that
there are not local analogs.  For instance, if the distribution of
observables is on average different at high redshift, as long as for
each high-redshift SN there is a similar low-redshift counterpart, the
peak brightness could, in principle, be correctly translated into an
accurate distance.  Within the local sample, there is no indication of
a correlation between host-galaxy metallicity and light-curve shape
\citep{Gallagher05}.

In the process of classifying and finding the redshifts for SNe from
the ESSENCE (Equation of State: SupErNovae trace Cosmic Expansion)
survey \citep{Miknaitis07, Wood-Vasey07}, we have obtained 107 spectra
which have accurate light-curve parameters such as $\Delta$ (a
light-curve width parameter), time of maximum light, and visual
extinction \citep{Matheson05, Foley08a}.  Most spectra in this sample
have low signal-to-noise ratios (S/N) compared to spectra of
low-redshift SNe.  This makes impractical a detailed analysis of each 
object individually to test for outliers.  However, by combining
the spectra to make composite spectra, we are able to study the mean
spectral properties of the samples.  

In Section~\ref{s:sample} we
discuss our low and high-redshift SN~Ia spectral samples.  We describe
our methods of creating composite spectra in Section~\ref{s:method}.
In Section~\ref{s:results} we present the composite spectra and
compare the two samples, while in Section~\ref{s:discussion} we
discuss the implications of these results.  We present our conclusions
in Section~\ref{s:conclusions}.  Throughout this paper we assume the
standard cosmological model with $(h, \Omega_{m}, \Omega_{\Lambda}) =
(0.7, 0.3, 0.7)$.

\section{Supernova Samples}\label{s:sample}

In order to test for potential evolution in SN~Ia spectra, we need to
explore the largest redshift range possible.  \citet{Riess07} have
obtained spectra of 10 objects with $z > 1$, with one SN~Ia at $z =
1.39$.  The large look-back time of these objects allows significant
time for progenitors to evolve between $z \approx 1$ and $z = 0$;
however, even at $z > 1$, many galaxies had already become metal rich
\citep[e.g.,][]{Cimatti04}.

In addition to these high-redshift SN spectra, through the first four
years of the ESSENCE campaign we have obtained 107 spectra of 88 
SNe~Ia with light curves that could be fit by MLCS2k2 \citep{Jha07}.  The
spectra were obtained by the Keck I and II 10~m telescopes, the Very
Large Telescope (VLT) 8~m, the Gemini North and South 8~m telescopes,
the Magellan Baade and Clay 6.5~m telescopes, the MMT 6.5~m telescope,
and the Tillinghast 1.5~m telescope at the F.~L. Whipple Observatory.
The spectra are analyzed individually by \citet{Matheson05} and
\citet{Foley08a}, while the light curves are presented by
\citet{Miknaitis07}.  The spectra are in the redshift range of $0.155
\leq z \leq 0.777$, with corresponding rest-frame phases of $-10.8 \leq
t \leq 20.9$~d relative to $B$-band maximum.  The objects span a wide
range of light-curve width from $-0.601 \leq \Delta \leq 0.843$,
corresponding to luminosities of $-19.7 \leq M_{V} \leq -18.6$ mag at
maximum brightness, respectively.

The observed wavelength ranges for the objects vary because the
spectra were obtained with different instruments and each spectrum has
been trimmed individually to remove the noisy ends of the spectrum.
The rest-wavelength ranges are even more disparate because of the
significantly different redshifts.  The rest-frame spectra span the
wavelength range 1940--8174~\AA.

For the purpose of comparison, we have also composed a sample of 397
spectra obtained mainly with the Kast double spectrograph
\citep{Miller93} on the Lick 3-m telescope \citep{Foley08b}, and 45
from \hstiue\, (to sample the UV portion of the spectrum; SNe
1980N, 1981B, 1982B, 1983G, 1986G, 1989B \citealt{Panagia07}, 1990N,
1991T, \citealt{Jeffery92}, 1992A \citealt{Kirshner93}, 2001eh, and
2001ep \citealt{Sauer08}).  The light curves of these objects are
presented elsewhere \citep{Buta85, Younger85, Ciatti88, Hamuy96,
Riess99:lc, Jha06, Ganeshalingam08}.  The Lick objects are observed
and reduced in a manner similar to the ESSENCE objects
\citep{Matheson01, Matheson05, Foley08a}.  These similarities reduce
the systematic differences between observing programs.  Although the
\hstiue\, spectra are not necessarily free of reduction errors,
the differences should be minor and these spectra have great utility
in our study.

\begin{deluxetable*}{lrrrrr}
\tablewidth{0pt}
\tablecaption{Number of Spectra For Redshift/Phase Binning\label{t:zage}}
\tablehead{
\colhead{Redshift} &
\colhead{$11 < t < -3$ d} &
\colhead{$-3 < t < 3$ d} &
\colhead{$3 < t < 10$ d} &
\colhead{$10 < t < 17$ d} &
\colhead{$17 < t < 23$ d}}

\startdata

0         & 53 & 55 & 50 & 46 & 35 \\
$0-0.2$   &  0 &  2 &  0 &  2 &  0 \\
$0.2-0.4$ &  7 & 10 & 10 &  8 &  3 \\
$0.4-0.6$ & 10 &  8 & 16 &  9 &  2 \\
$0.6-0.8$ &  1 &  7 &  8 &  2 &  0 \\

\enddata

\end{deluxetable*}

\begin{deluxetable}{lrrr}
\tablewidth{0pt}
\tablecaption{Number of Spectra For Redshift/$\Delta$ Binning\label{t:zdelta}}
\tablehead{
\colhead{Redshift} &
\colhead{$\Delta < -0.15$} &
\colhead{$-0.15 < \Delta < 0.3$} &
\colhead{$\Delta > 0.3$}}

\startdata

0         & 141 & 158 & 89 \\
$0-0.2$   &   2 &   2 &  0 \\
$0.2-0.4$ &  15 &  19 &  4 \\
$0.4-0.6$ &  26 &  21 &  0 \\
$0.6-0.8$ &  15 &   3 &  0 \\

\enddata

\end{deluxetable}

\begin{figure}
\epsscale{0.6}
\rotatebox{90}{
\plotone{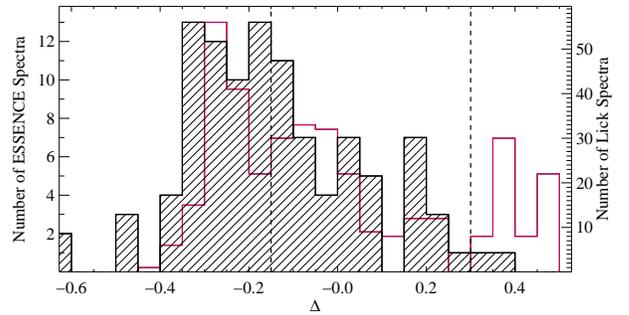}}
\caption{A histogram of the number of spectra per $\Delta$ bin.  The
thick line with shading represents the ESSENCE objects, while the thin
line represents the Lick objects.  The dotted lines represent the
distinction between the luminosity subclasses of SNe~Ia (overluminous,
normal, and underluminous correspond to $\Delta < -0.15$, $-0.15 <
\Delta < 0.3$, and $\Delta > 0.3$, respectively) from
\citet{Jha06}.}\label{f:delta_hist}
\end{figure}

\begin{figure}
\epsscale{0.6}
\rotatebox{90}{
\plotone{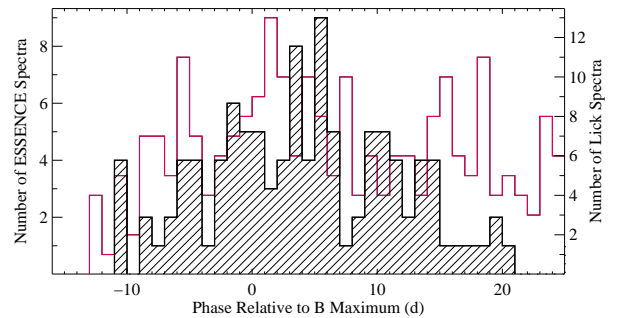}}
\caption{A histogram of the number of spectra per phase bin relative
to the $B$-band maximum bin.  The thick line with shading represents the
ESSENCE objects, while the thin line represents the Lick
objects.}\label{f:age_hist}
\end{figure}

\begin{figure}
\epsscale{0.6}
\rotatebox{90}{
\plotone{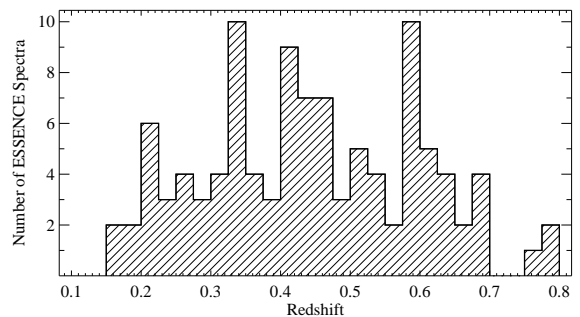}}
\caption{A histogram of the number of ESSENCE spectra per redshift
bin.}\label{f:redshift_hist}
\end{figure}

In Figures~\ref{f:delta_hist}, \ref{f:age_hist}, and
\ref{f:redshift_hist}, we present histograms showing the $\Delta$,
phase, and redshift distributions, respectively.  The ESSENCE and Lick
samples have similar distributions, and more importantly, the Lick
spectra sample the entire range of phases and almost the entire
$\Delta$ range of the ESSENCE sample, allowing for an unbiased
comparison.  The total numbers of spectra in various redshift, phase,
and $\Delta$ bins are presented in Tables~\ref{t:zage} and
\ref{t:zdelta}.  All composite spectra presented in this paper are
publicly
available\footnote{\url{http://astro.berkeley.edu/$\sim$rfoley/composite/}}.

\section{Method}\label{s:method}

  \subsection{Reprocessing}\label{ss:red_gal}

To properly combine our sample of SN spectra, we must first attempt to
remove any effects of non-SN sources from each spectrum.  The two main
effects are reddening by host-galaxy dust and galaxy-light
contamination.

The reddening of the spectra causes the spectral shape and relative
feature strengths to change.  To correct for the distortion introduced
by reddening, we have used the MLCS reddening parameters listed in
Table~\ref{t:hiz} and a standard reddening law \citep{Cardelli89} to
deredden our spectra. Since the ESSENCE photometry consists of only
two bands, it is not possible to independently determine both $A_{V}$
and $R_{V}$. To that extent, our corrected spectra may not be
properly dereddened.  The relative flux difference between a spectrum
with $A_{V} = 0.3$ mag with $R_{V} = 2.1$ and $R_{V} = 3.1$ is less than
10\% for wavelengths longer than 3400~\AA, and it is possible that the
reddening parameters account for some intrinsic color variability
within the SN sample \citep{Conley07}.  However, we have performed our
analysis with both a low-extinction sample ($A_{V} < 0.3$ mag ) and
our complete, treated sample (see Section~\ref{ss:extinction}),
yielding similar results.  Therefore, the errors in $A_{V}$ and
$R_{V}$ are not dominant.

At high redshift, the size of the SN host galaxy and the average
offset between the host galaxy and the SN are small relative to the
slits of our spectrographs.  Consequently, it is often difficult to
completely remove the galaxy-light contribution from the total light
at the position of the SN.  For some ESSENCE spectra, we employ a
deconvolution technique \citep{Blondin05}, which separates the SN and
galaxy light better than simple background subtraction.  With several
bands of photometry for the galaxy, one can properly model the galaxy
type, allowing one to find a matching template spectrum to subtract
off the SN spectrum \citep{Howell05}.  However, using ESSENCE's current
photometry this is not possible.  A campaign in underway to image the
ESSENCE fields in \bvz\!\!\!, and the resulting data should allow one to
properly model the galaxy spectral-energy distribution (SED).

Obtaining a spectrum of the host galaxy after the SN has faded and
subtracting an appropriate percentage of the host-galaxy spectrum from
the SN spectrum is the best method to remove host-galaxy light
contamination.  This method, however, is typically very
time intensive, and has not been performed on our sample.  We have a
program underway to obtain the host-galaxy spectra; however,
observations have not yet begun.  In this paper, we neglect the
galaxy-light contamination in our spectra when making the composite
spectra and then choose to compare our samples in a way that accounts
for it (see Section~\ref{s:results}).

%


\subsection{Construction of the Composite Spectra}

The composite spectra were constructed by first deredshifting the
individual spectra to the rest frame.  The redshifts have errors up to
$0.01$ which might artificially widen some spectral features in the
composite spectrum.  If we wish to look at a dereddened spectrum, we
then deredden the spectra based on our values of $A_{V}$ and $R_{V}$
found from fitting templates to the light curves ($R_{V}$ is fixed at
3.1 for all ESSENCE objects, and unless the fits require $R_{V} \ne
3.1$, the low-redshift objects also have $R_{V}$ fixed to 3.1;
\citealt{Jha07}).  We then put the spectra through a low-pass filter
to remove residual night-sky lines, cosmic rays, and host-galaxy
absorption and emission lines.  By comparing the original spectra to
a heavily smoothed version of each spectrum, we then determine the
pixel-by-pixel S/N.  Finally, we average the spectra, weighting by
S/N.  Since not all spectra have a common wavelength range
(particularly comparing the \iue\, spectra to the Lick spectra), we
first constructed a composite spectrum from the majority of spectra
with overlapping wavelengths, and then used that temporary composite
spectrum (which overlapped with all spectra) to match the fluxes of
the individual spectra.


There are some intrinsic issues to constructing a composite SN
spectrum.  Despite having hundreds of SN spectra, to ensure a sufficient
number of spectra per parameter bin we must still have somewhat large
bins for some parameters such as phase.  This can smear out certain 
features. However, since SN spectra tend to evolve smoothly over small time
periods, averaging over a phase bin is a reasonable estimate for the
average phase spectrum.  Other issues are intrinsic differences in SN
spectra.  For a given phase and $\Delta$, spectra still differ from one
object to another \citep{Matheson07}.  Or, within a particular bin,
objects can have significantly different expansion velocities or
colors.  When constructing a composite spectrum, most of these
spectral differences do not change the composite spectrum from looking
like the ``true'' average SN spectrum.  However, differences in
expansion velocity will tend to make spectral features wider and
shallower.  Since we are comparing composite spectra to each other and
not a composite spectrum to an individual spectrum, as long as the
underlying samples are similar, the effects of differing expansion
velocities will not create significant differences in the composite spectra.

The ESSENCE SNe come from a blind search, resulting in many objects in
low-luminosity host galaxies.  The Lick sample, on the other hand,
comes primarily from targeted searches and is biased to
higher-luminosity host galaxies.  This may result in slightly
different samples, however the similar $\Delta$ distributions for the
ESSENCE and Lick samples suggest against this.  However, in creating
composite spectra, since the spectra are weighted by S/N, the
composite spectra will be weighted more towards intrinsically
overluminous SNe~Ia.  Since this effect occurs in both samples, this
should not affect our analysis.

  \subsection{Determining Spectral Variance}

We want to determine both the average spectrum of a given phase,
$\Delta$, and redshift, as well as the variation about that average
spectrum.  To do this, we implement a boot-strap sampling (with
replacement) algorithm to estimate the variance \citep{Efron82}. The
variance is a combination of the noise in our spectra, any systematic
effects during the reduction process (such as poor sky-line removal
and incomplete galaxy-light subtraction), and the inherent variance in
the SN sample.


For the Lick sample, the spectra are generally of very high quality
with little noise or systematic issues.  As such, the variance in
these spectra is dominated by the intrinsic scatter among the objects.  
A detailed analysis of these spectra will be presented in a future
paper \citep[for another low-redshift composite spectrum derived
from eight objects, see][]{James06}.  For the ESSENCE sample, the
Poisson noise dominates at the bluest wavelengths since there are
fewer spectra adding to the composite in this wavelength region and 
the spectra have lower S/N in the UV.  The reddest wavelengths are 
dominated by reduction issues, specifically poor sky subtraction
residuals.  The observed wavelengths corresponding to night-sky lines
are weighted less, but the average noise over large wavelength ranges
remains higher at observed near-IR wavelengths than at visual
wavelengths.  The other major factor at long wavelengths is the large
variation in host-galaxy contamination.  Since the SEDs will vary
dramatically based on the amount of galaxy-light contamination (SNe~Ia
are blue relative to galaxies), there is additional variance unrelated
to the intrinsic variability of SN spectra.

\section{Results}\label{s:results}

  \subsection{Comparison of Low-Redshift Templates}\label{ss:nugent}

To understand the subtle differences between the ESSENCE and
low-redshift composite spectra, we must first examine the differences
between different low-redshift composite spectra.

The original Nugent template spectra \citep{Nugent02} were constructed
from 84 spectra (31 for the maximum-light template) of which 63\%
(52\% near maximum light) come from SNe 1989B, 1992A, and 1994D.  Two
of these objects have been considered slightly atypical SNe~Ia, with
strong dust absorption \citep[SN 1989B; ][]{Wells94} or anomalous
luminosity and colors \citep[SN 1994D; ][]{Richmond95, Patat96}.
Since no SN is the perfect example of a SN~Ia, the presence of these
objects in the composite is not worrisome.  However, having the
majority of the sample rely on a few objects can severely skew the
composite away from a true average.  The Nugent templates have since
been
updated\footnote{http://supernova.lbl.gov/$\sim$nugent/nugent\_templates.html .},
with more objects.  However, the influence of these atypical SNe~Ia is still
strong.

A new spectral template has been constructed by \citet{Hsiao07} for
the Supernova Legacy Survey (SNLS).  This template is created by
combining low and high-redshift spectroscopy and photometry.  It is
created from a more varied sample than the Nugent template, but still
has a very limited low-redshift sample.

\begin{figure}
\epsscale{0.9}
\rotatebox{90}{
\plotone{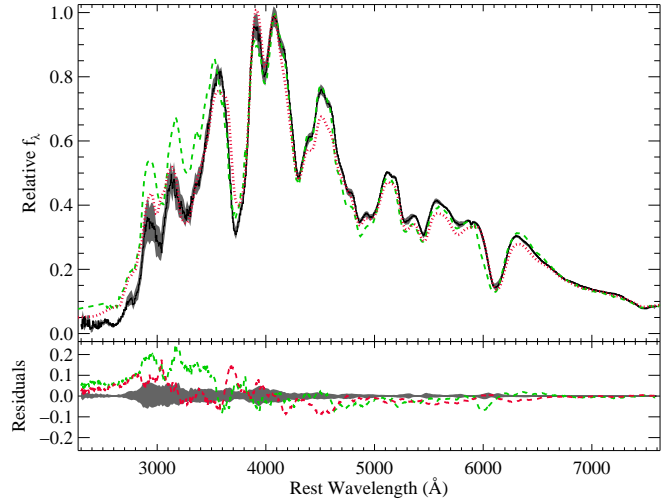}}
\caption{({\it top panel}):  Composite low-$z$ maximum-light SN~Ia
spectra.  The black curve is the composite spectrum of the Lick
sample; the green (dashed) curve is the Nugent template spectrum; and
the red (dotted) curve is the SNLS template spectrum.  The grey region
is the $1\sigma$ boot-strap variation of the Lick composite spectrum.
({\it bottom panel}):  The difference between the Lick composite spectrum
and the Nugent and SNLS template spectra.  The grey region is the
$1\sigma$ boot-strap variation of the Lick composite
spectrum.}\label{f:lick_nug}
\end{figure}

The Lick maximum-light composite spectrum, the Nugent maximum-light
template spectrum, and the new SNLS template spectrum are shown in 
Figure~\ref{f:lick_nug}.  The Nugent template is bluer at this phase, 
particularly in the UV and the near-UV.  If normalized at 4000~\AA, 
the Lick composite spectrum has higher flux levels than either the 
SNLS or Nugent template for wavelengths somewhat redder than 4300~\AA, 
although it is very similar to the Nugent template in this region and 
to both templates for wavelengths redder than 4700~\AA.  Besides the 
feature at 4500 \AA, the SNLS template is very
similar to the Lick composite.  

The only other region with major differences is in the UV 
(where the Nugent template also deviates from
the Lick composite spectrum).  Both the Nugent and SNLS templates are
constructed by warping the spectra to match the colors of a normal
SN~Ia at the phase of the spectrum (for the SNLS template, this is
done by using light-curve information in the construction of the
template).  However, the large dispersion in the $U$ band
\citep{Jha06} along with extrapolation into the UV makes the Nugent
template UV continuum dubious.  The Lick spectra were all observed and
reduced in a consistent manner, producing relative spectrophotometry
accurate to \about$\pm5$\% \citep{Matheson00:93j}.  The colors for the
three spectra as well as the colors of the MLCS $\Delta = 0$ template
light curves are presented in Table~\ref{t:temp_colors}.  

Considering the relative diversity of the Lick and Nugent samples and the 
methods of producing the composite spectra, we consider the Lick composite
spectrum to be more reliable than the Nugent template.  The SNLS
template uses high-redshift SN data (both light curves and spectra) in
the construction of their template.  As a result, this template should
not be used for comparison to high-redshift SNe~Ia.  In particular, the
UV portion of the SNLS template is heavily weighted to high-redshift
SNe~Ia.  Most importantly, the spectra in the Lick and ESSENCE samples
were reduced and the composite spectra were constructed in the same
manner.  For our purposes, the Lick composite spectrum is superior to
the alternatives and will be used for comparison in the rest of this
paper.

\begin{deluxetable}{lrrrr}
\tablewidth{0pt}
\tablecaption{Low-Redshift Maximum-Light Colors\label{t:temp_colors}}
\tablehead{
\colhead{Spectrum} &
\colhead{$U-B$} &
\colhead{$B-V$} &
\colhead{$V-R$} &
\colhead{$V-I$} \\
\colhead{} &
\colhead{(mag)} &
\colhead{(mag)} &
\colhead{(mag)} &
\colhead{(mag)}}

\startdata

MLCS $\Delta = 0$ & $-0.47$ & $-0.07$ &   0.00  & $-0.29$ \\
Lick              & $-0.33$ & $-0.04$ & $-0.04$ & $-0.34$ \\
Nugent            & $-0.45$ & $-0.07$ &   0.00  & $-0.35$ \\
SNLS              & $-0.41$ & $-0.06$ & $-0.02$ & $-0.26$ \\

\enddata

\end{deluxetable}

The relative strengths of most absorption lines in the Nugent, SNLS, and 
Lick spectra are similar.  The line velocities are also similar, with the
exception of \ion{Si}{2} 6355, which has a blueshift at maximum light
of $-12,300$ \kms (Nugent template) versus $-11,200$ \kms (SNLS
template) and $-11,400$ \kms (Lick composite).  Another interesting
difference is the weaker feature at \about 3000~\AA\ in the SNLS
template compared to the Nugent template and Lick composite spectrum.
Although \citet{Hsiao07} show no major difference at this wavelength
when including or excluding high-redshift spectra, the final SNLS
template contains both high-redshift spectra and light-curve
information, which may result in a bias for the \about 3000~\AA\
feature (see Section~\ref{ss:lines}).

  \subsection{Maximum-Light Spectrum}\label{ss:maxlight}

The nature of the ESSENCE search dictates that most objects are
observed spectroscopically near maximum light.  As seen in
Figure~\ref{f:age_hist}, most spectra were obtained near maximum
light (64/107 have rest-frame phases within one week of maximum
light), with a median of 4.1~d after maximum light.  Since objects are
brightest at maximum light, the highest S/N spectra are usually
obtained at this phase and the contrast between the SN and the
underlying galaxy is largest.  These factors together make the
maximum-light composite spectrum a higher S/N spectrum.

\begin{figure}
\epsscale{0.9}
\rotatebox{90}{
\plotone{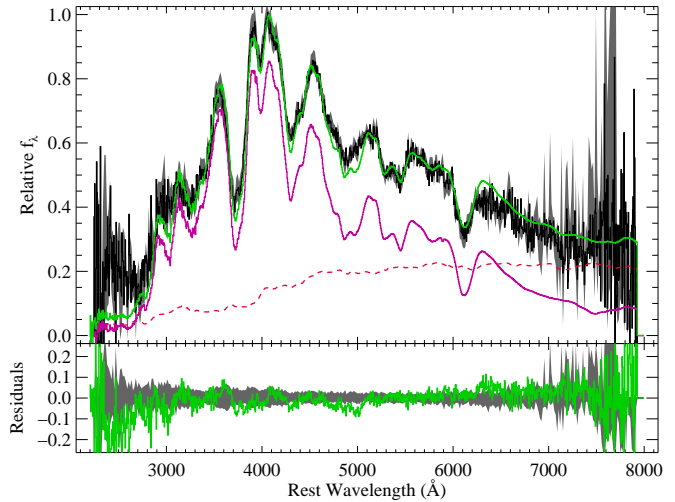}}
\caption{({\it top panel}): ESSENCE maximum-light composite SN~Ia
spectrum.  The black, relatively noisy curve is the ESSENCE composite
spectrum.  The purple curve, slightly below the black curve, is the
Lick maximum-light composite spectrum.  The red (dashed) curve is an Sb
galaxy spectrum. The green curve is the addition of the red and purple
curves, fit to match the ESSENCE composite spectrum.  The combination
is 35\% of host-galaxy spectrum in the region 4000--6000~\AA.  The grey
region is the $1\sigma$ boot-strap variation of the ESSENCE composite
spectrum.  The large variance at red wavelengths is due to strong
sky-subtraction residuals in individual spectra and not intrinsic
variability in the SN sample.  ({\it bottom panel}):  The difference
between the ESSENCE and Lick composite spectra.  The grey region is the
$1\sigma$ boot-strap variation of the ESSENCE composite
spectrum.}\label{f:essmax_comp}
\end{figure}

In Figure~\ref{f:essmax_comp}, we present the ESSENCE maximum-light
composite spectrum with average parameters $\mean{z} = 0.37$,
$\mean{t} = -0.8$~d, and $\mean{\Delta} = 0.01$.  Comparing the
spectrum to the Lick maximum-light spectrum (which has average
parameters $\mean{z} = 0.02$, $\mean{t} = 0.5$~d, and $\mean{\Delta} =
-0.02$), it is obvious that the two spectra have rather different
colors, with the ESSENCE spectrum being redder than the Lick spectrum.

As described in Section~\ref{ss:redshift}, the large projected size of
the slit for high-redshift objects makes isolating the SN from the
host galaxy difficult.  It is much easier to separate the SN from the
galaxy light at low redshift.  Our ability to remove host-galaxy light 
from any given SN spectrum hinges on how well we can model the galaxy
background, which is highly dependent on how isolated the SN is from
its host galaxy.  To account for the difference in the amount of
galaxy light remaining in a SN spectrum, we fit a combination of a
galaxy spectrum and the Lick composite spectrum to match the ESSENCE
spectrum.  Similar to the method of \citep{Howell05}, using five
galaxy spectra templates (E, S0, Sa, Sb, and Sc) and varying the
galaxy light from 0\% to 100\% of the comparison spectrum, we find a
best-fit combination.  As seen in Figure~\ref{f:essmax_comp}, by
adding some galaxy light (35\% of an Sb galaxy spectrum for the
maximum-light spectrum) to the Lick spectrum, we can match the Lick
and ESSENCE spectra quite well.  We believe galaxy-light contamination
to be the main factor in the discrepancy between the Lick and ESSENCE
continua.

Properly determining the galaxy contamination is crucial in determining
differences between our two samples.  We have performed several tests
to determine the validity of our claims with different galaxy
contamination.

In addition to fitting template galaxy spectra to our observed spectra, we have
reconstructed the best-fit galaxy spectra from galaxy eigenspectra.
Using the first four Sloan Digital Sky Survey (SDSS) galaxy eigenspectra \citep{Yip04},  we fit
the eigenspectra to the residuals of the ESSENCE and Lick spectra.
Higher-order eigenspectra are dominated by emission lines and
high-frequency modulations.

The process consists of reducing the $\chi^{2}$ of
\begin{equation}
  f_{\rm ESSENCE} - \left ( a f_{\rm Lick} + \mathbf{b}
    \mathbf{U}^{t}_{\rm SDSS} \right ),
\end{equation}
\noindent
where $f_{\rm ESSENCE}$ is the flux vector of the ESSENCE composite
spectrum, $f_{\rm Lick}$ is the flux vector of the Lick composite
spectrum, $\mathbf{U}_{\rm SDSS} = \{e_{i}\}$, $1 \le i \le 4$ is the
matrix of SDSS eigenspectra, $a$ is a free parameter, and $\mathbf{b}$
is the eigenvector that best describes the $f_{\rm ESSENCE} - a f_{\rm
Lick}$ residual.  The best-fit reconstructed galaxy spectrum is
comprised of 83.5\%, 7.2\%, 5.6\%, and 3.7\% of the first-four
eigenspectra, respectively.  The first eigenspectrum is the average of
the SDSS galaxy sample and resembles an Sb galaxy.

\begin{figure}
\epsscale{0.9}
\rotatebox{90}{
\plotone{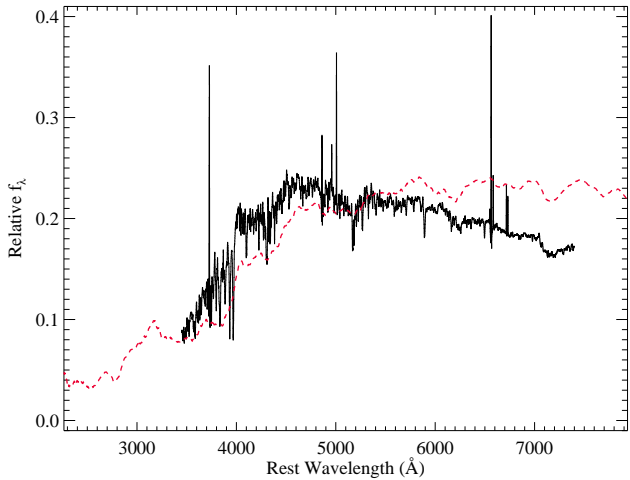}}
\caption{The best-fit PCA-reconstructed and template galaxy spectra
(black and red curves, respectively).  Since the amount of galaxy
light in the fit changes based on the SED of the galaxy spectrum, the
scaling between the spectra is arbitrary.}\label{f:gal}
\end{figure}

In Figure~\ref{f:gal}, we present the principal-components analysis
(PCA) reconstructed best fit, and our Sb galaxy template spectrum.
The PCA-reconstructed and template spectra are fit with differing
amounts of SN light, so their absolute scaling is approximately the
same, but slightly different.  Despite this, we can still see that the
PCA and Sb template spectra have very similar SEDs, differing the most
around 4200~\AA\ and at wavelengths longer than 6500~\AA.
Extrapolating the PCA spectrum, it appears that the spectra do not
differ drastically in the UV.

\begin{figure}
\epsscale{0.9}
\rotatebox{90}{
\plotone{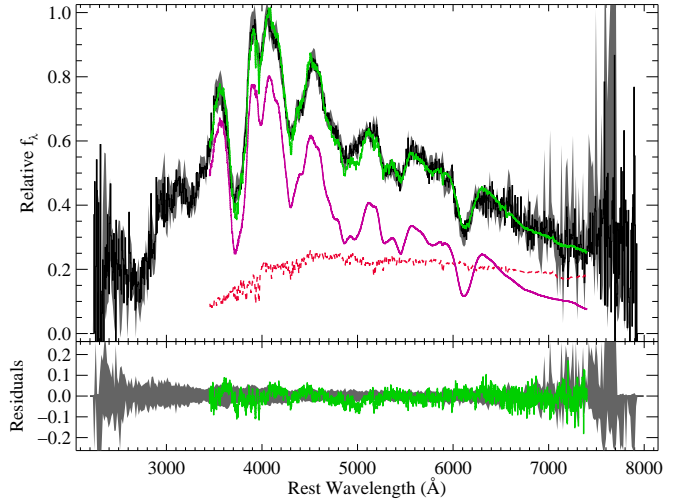}}
\caption{({\it top panel}): ESSENCE maximum-light composite SN~Ia
spectrum.  The black, relatively noisy curve is the ESSENCE composite
spectrum.  The purple curve, slightly below the black curve, is the
Lick maximum-light composite spectrum.  The red (dashed) curve is the
sigma-clipped SDSS PCA-reconstructed galaxy spectrum. The green curve
is the addition of the red and purple curves, fit to match the ESSENCE
composite spectrum.  The combination is 40\% of galaxy spectrum in the
region 4000--6000~\AA.  The grey region is the $1\sigma$ boot-strap
variation of the ESSENCE composite spectrum.  The large variance at
red wavelengths is due to strong sky-subtraction residuals in
individual spectra and not to intrinsic variability in the SN sample.
({\it bottom panel}):  The difference between the ESSENCE and Lick
composite spectra.  The grey region is the $1\sigma$ boot-strap
variation of the ESSENCE composite spectrum.}\label{f:essmax_pca}
\end{figure}

Figure~\ref{f:essmax_pca} shows the fit and residuals of the
PCA-reconstructed galaxy spectrum (similar to
Figure~\ref{f:essmax_comp}).  The residuals for the PCA-reconstructed
fit are smaller than those for the template fit.  Since the addition
of the PCA-reconstructed galaxy spectrum to the Lick composite
spectrum gives, by construction, the smallest deviation from the
ESSENCE composite spectrum, we consider any additional differences to
come from sources other than galaxy contamination.  Considering the
overall similar results and smaller wavelength range of the
PCA-reconstructed galaxy spectrum, we will use the template galaxy
spectrum for comparisons, but note differences with the
PCA-reconstructed galaxy spectrum when appropriate.

It is unlikely that miscalculated reddening corrections can cause the
differences in the continua.  Using the unreddened sample of spectra
produces a similar continuum (see Section~\ref{ss:extinction}).
Although it is possible that SN evolution could cause the discrepancy,
no current models predict that the spectrum would change in this way.


After correcting for galaxy-light contamination, by adding galaxy
light to the Lick composite spectrum\footnote{Throughout this paper, 
in all comparisons, ESSENCE composite spectra are presented unchanged, 
and Lick composite spectra have additional galaxy light to match the
ESSENCE spectra, unless specifically noted.}, the Lick spectrum is very
similar to the ESSENCE composite spectrum.  Despite the similarities,
there are two differences worth noting: the ESSENCE composite spectrum
lacks the absorption at \about 3000~\AA\ and has a weaker feature at
\about 4900~\AA.  The feature at \about 3000~\AA\ is attributed to
multiple \ion{Fe}{2} lines \citep{Branch86}, while the feature at
\about 4900~\AA\ is attributed to \ion{Si}{2}, \ion{Fe}{2}, and
\ion{Fe}{3}, with the red wing, where the discrepancy occurs, being
\ion{Fe}{3} $\lambda 5129$.  We will use the convention of
\citet{Garavini07} and call the entire feature ``\ion{Fe}{2} 4800,''
naming the lines individually only when we are discussing particular
species contributing to the feature.

Using the prescription of \citet{Garavini07}, we measure the
pseudo-equivalent widths (pEWs) of several features in both the
ESSENCE and Lick composite spectra.  These values are presented in
Table~\ref{t:ew}.  We note that because of the host-galaxy
contamination, the pEWs of the ESSENCE composite spectrum are lower
limits.  We have calculated the pEWs of the Lick composite spectrum
both with and without galaxy light added.  The values of the spectrum
with galaxy light added should be compared directly to the values of
the ESSENCE spectrum; however, the values without galaxy light give
both an accurate measurement and an upper limit for the pEWs.  There
is significant systematic uncertainty to these measurements from the
placement of the continuum.  We did not attempt to model the exact
systematic errors, but they are typically \about 10~\AA\
\citep{Garavini07}.

\begin{deluxetable*}{lcccccc}
\tablewidth{0pt}
\tablecaption{Pseudo-Equivalent Widths\label{t:ew}}
\tablehead{
\colhead{Spectrum} &
\colhead{\ion{Ca}{2} H\&K} &
\colhead{\ion{Si}{2} 4000} &
\colhead{\ion{Mg}{2} 4300} &
\colhead{\ion{Fe}{2} 4800} &
\colhead{\ion{S}{2} W} &
\colhead{\ion{Si}{2} 6150} \\
\colhead{} &
\colhead{(\AA)} &
\colhead{(\AA)} &
\colhead{(\AA)} &
\colhead{(\AA)} &
\colhead{(\AA)} &
\colhead{(\AA)}}

\startdata

Lick          & 112.8 (0.6) & 12.4 (0.6) & 86.4 (0.4) & 140.9 (0.4) & 68.5 (0.4) & 99.5 (0.6) \\
Lick + Galaxy & 100.7 (0.5) &  9.6 (0.4) & 73.6 (0.3) &  98.8 (0.2) & 43.4 (0.2) & 61.4 (0.3) \\
ESSENCE       &  99.7 (1.1) &  9.7 (0.5) & 73.8 (0.8) &  78.5 (0.8) & 45.7 (0.8) & 51.7 (1.7) \\

\enddata

\end{deluxetable*}

For most features, the ESSENCE and Lick composite spectra have similar
pEWs.  \citet{Bronder07} found a possible difference in the strength
of the \ion{Mg}{2} 4300 feature, but we see no difference in our
composite spectra.  We will show a full analysis similar to the
\citet{Bronder07} study in \citet{Foley08a}.  The \ion{Si}{2} 6150 pEW
has a slightly smaller value in the ESSENCE spectrum.  Similarly, the
\ion{Fe}{2} 4800 feature is much weaker in the ESSENCE spectrum, which
is likely because of a weaker \ion{Fe}{3} $\lambda 5129$ line.  Using
the PCA-reconstructed galaxy spectrum, we see that the Lick
\ion{Si}{2} feature has a similar pEW to the ESSENCE composite
spectrum.  The \ion{Fe}{2} 4800 feature continues to have a larger
(although slightly smaller than using the galaxy template spectrum,
$\text{pEW} = 94.9 \pm 0.2$ \AA) pEW than the ESSENCE composite
spectrum.  The PCA-reconstructed galaxy spectrum does not cover the
3000~\AA\ region.

  \subsection{Low-Extinction Spectrum}\label{ss:extinction}

\begin{figure}
\epsscale{0.9}
\rotatebox{90}{
\plotone{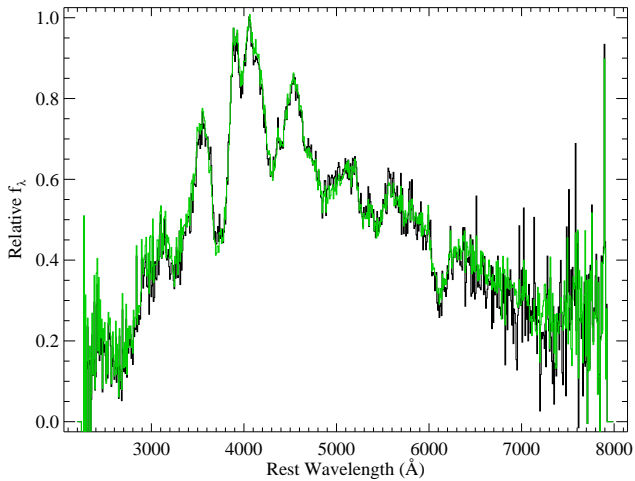}}
\caption{ESSENCE low-$A_{V}$ maximum-light composite SN~Ia spectrum.
The black curve is the ESSENCE low-$A_{V}$ composite spectrum and the
green curve is the ESSENCE full-sample maximum-light composite
spectrum (as shown in Figure~\ref{f:essmax_comp}).  The spectra
contributing to the full-sample spectrum have been dereddened by the
values listed in Table~\ref{t:hiz}.  The low-$A_{V}$ spectrum has not
been dereddened.  The low-$A_{V}$ and full-sample spectra are nearly
identical, indicating that the difference between the Lick and ESSENCE
spectra is not the result of miscalculated reddening
corrections.}\label{f:av_comp}
\end{figure}

As mentioned in Section~\ref{ss:maxlight}, a possible explanation for
the difference between the Lick and ESSENCE composite spectra is a
miscalculation of the reddening from the light curves.  To test this
hypothesis, we have created an un-dereddened composite spectrum using
only objects with $A_{V} < 0.3$ mag.  This subsample is less dependent
on the precise values of $A_{V}$ and $R_{V}$, reducing those potential
sources of systematic error. Figure~\ref{f:av_comp} shows both the
full, dereddened and low-$A_{V}$ (with no dereddening) ESSENCE
composite spectra.  Both the low-$A_{V}$ and full, dereddened ESSENCE
samples produce nearly identical spectra.  There is a slight
difference between the continua, with the low-$A_{V}$ sample being
slightly redder.  This is probably due to the slight extinction that
has not been corrected in the contributing spectra.

  \subsection{Redshift-Binned Spectra}\label{ss:redshift}

\begin{figure}
\epsscale{0.9}
\rotatebox{90}{
\plotone{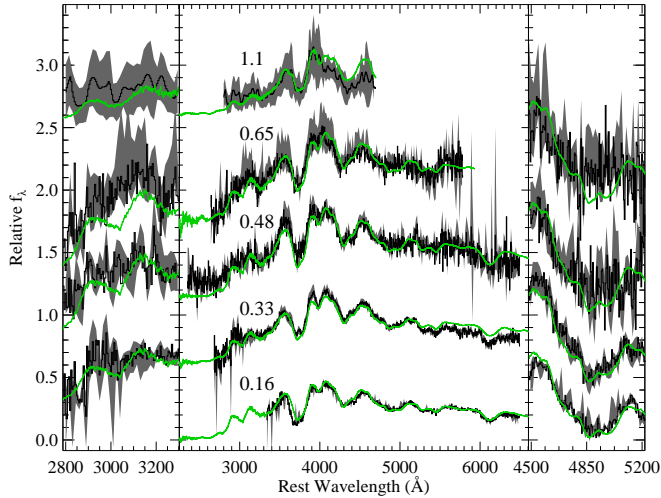}}
\caption{ESSENCE maximum-light composite SN~Ia spectra for different
redshift bins.  The spectra have average redshifts of 0.16, 0.33,
0.48, and 0.64, as well as average phases of $-2.2$ d, $-0.6$ d, $-0.1$
d, and $-0.1$ d, respectively.  The grey regions are the $1\sigma$
boot-strap variation.  The green curves are the Lick maximum-light
composite spectrum with 34\%, 35\%, 41\%, and 49\% galaxy light for the
comparisons to the $\mean{z} = 0.16$, $\mean{z} = 0.33$, $\mean{z} =
0.48$, and $\mean{z} = 0.65$ bins, respectively.  The \hst\, composite
spectrum from \citet{Riess07}, with $\mean{z} = 1.1$, is composed of
individual spectra spanning phases of 0--10~d past maximum
brightness.  For comparison, a Lick composite spectrum was constructed
from low-redshift spectra with similar phases.  For the \hst\, composite
spectrum, no galaxy light was added to the Lick comparison
spectrum.  We argue that galaxy-light contamination should be lower in
the \hst\, spectra (see text), and this is confirmed by our
galaxy-light fitting routine which gives a best fit with no galaxy
light.  The left and right panels show the regions surrounding the
\about 3000~\AA\ and \ion{Fe}{2} 4800
features.}\label{f:redshift_comp}
\end{figure}

We have divided the full sample of ESSENCE spectra into four redshift
bins with $\Delta z = 0.2$.  Since the ESSENCE sample is confined to
$z < 1$, we also utilize the composite spectrum of \citet{Riess07},
which has $\mean{z} = 1.1$.  The composite spectra with mean phases
$\mean{t} = -2.2$, $-0.6$, $-0.1$, and $-0.1$~d for the $0 < z < 0.2$,
$0.2 < z < 0.4$, $0.4 < z < 0.6$, and $0.6 < z < 0.8$ bins
(respectively) are presented in Figure~\ref{f:redshift_comp}.  Broadly,
the spectra resemble each other with no major differences except for
redder continua with increasing redshift.  We attribute the redder
colors with increasing redshift to an increase in the galaxy fraction
of the spectrum.

\begin{figure}
\epsscale{1.1}
\rotatebox{90}{
\plotone{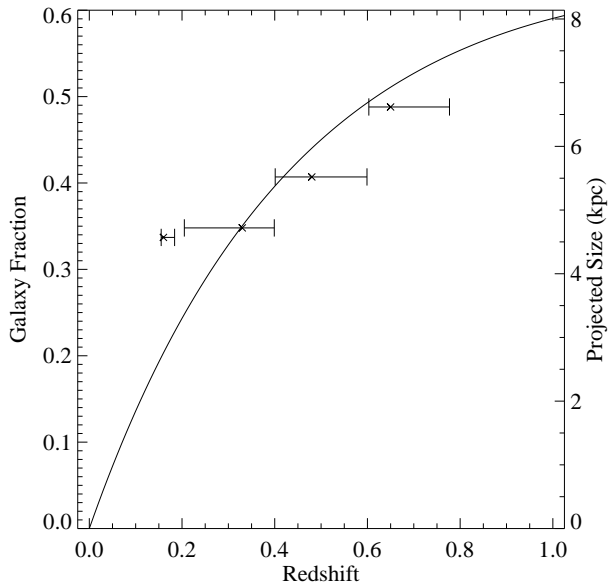}}
\caption{The galaxy fraction for ESSENCE maximum-light composite
spectra of varying redshifts found by fitting the Lick maximum-light
composite spectrum and a varying amount of galaxy light to the ESSENCE
spectra.  The solid curve represents the projected size of 1\arcsec\
with redshift.  The typical size of a slit used to obtain the spectra
is 1\arcsec.}\label{f:galfrac}
\end{figure}

Using the method described in Section~\ref{ss:maxlight}, we determined
the galaxy-light fraction of each redshift-binned composite spectrum.
Figure~\ref{f:galfrac} shows the galaxy fraction of the
redshift-binned ESSENCE composite spectra.  For comparison, we also
show the relationship between the projected distance of 1\arcsec\ and
redshift, given by
\begin{equation}
  d_{\perp} = 4.848 \times d_{L} \left ( \frac{\theta}{1\arcsec}
                \right ) (1 + z)^{2} \text{ kpc},
\end{equation}
\noindent
where $d_{L}$ is the luminosity distance given in Gpc.  If the
separation between the SN and the galaxy is less than the larger of
the slit width and the seeing, both typically 1\arcsec, there will be
galaxy contamination in the SN spectrum.  If the galaxy is larger than
the seeing, the separation can be larger and the SN spectrum can still
be contaminated by galaxy light.  If the physical distance between SN
and galaxy nucleus, galaxy luminosity, and galaxy size do not change,
either by galaxy evolution or selection effects, we expect the galaxy
fraction to increase with $d_{\perp}$ since larger redshift
corresponds to more of the galaxy in the slit.  Figure~\ref{f:galfrac}
shows this trend, further suggesting that the main difference between
the Lick and ESSENCE composite spectra is the result of galaxy-light
contamination.  The \hst\, composite spectrum is not affected by
galaxy-light contamination since the high angular resolution and
narrow slits of \hst\, give comparable conditions to those of the
low-redshift objects.  Confirming this hypothesis, our galaxy-light
fitting routine yields a best fit of 0\% galaxy light for the \hst\,
composite spectrum.


In the three lowest-redshift bins, we are able to examine the
\ion{Si}{2} $\lambda 6355$ line.  In these bins, it does not appear to
be stronger at higher redshift, consistent with that found for the
maximum-light spectrum (see Section~\ref{ss:maxlight} and
Table~\ref{t:ew}).  The major differences between the total
maximum-light composite ESSENCE spectrum and the Lick composite
spectrum are the lack of absorption at \about 3000~\AA\ in the ESSENCE
spectrum and the weaker \ion{Fe}{3} $\lambda 5129$ line.  Both lines
are shown in detail in Figure~\ref{f:redshift_comp}.  The \about 3000~\AA\ 
line is only available in the three highest-redshift bins.  In
the $0.2 < z < 0.4$ bin, there appears to be some absorption at this
wavelength.  The $0.4 < z < 0.6$ bin does not appear to have any
absorption, but the line does have a positive pEW that is
$3.3\sigma$ from zero.  For the highest-redshift bin ($z > 0.6$), the
absorption is lacking.  The \ion{Fe}{3} $\lambda 5129$ line is within
the wavelength range of all ESSENCE redshift-binned composite spectra.
The lowest two redshift bins appear to have line strengths consistent
with the Lick composite spectrum.  However, the highest two redshift
bins appear to be lacking some absorption.

The UV portion of the Lick composite spectrum consists of only a few
spectra from even fewer objects.  As such, the \about 3000~\AA\
feature might be anomalous for typical SNe~Ia.  However, the presence
of this line in one of our ESSENCE spectra but not in the two
highest-redshift bins suggests that the line has some redshift
dependence.  Since we preferentially select objects that have bright
UV fluxes at higher redshift, this may be a bias in the sample.  The
$\mean{z} = 1.1$ composite spectrum from \citet{Riess07} does show a
slight dip at 3000~\AA, but the spectrum is rather noisy at $\lambda <
3500$~\AA\ and the dip is narrow compared to other features, so it is
likely insignificant.  If the line is present in the \hst\, spectrum but
not in the highest-redshift bins of the ESSENCE sample, then it is
likely associated with another property that changes with redshift in
the ESSENCE sample, but is unaffected in the \hst\, objects, presumably
the result of a selection effect.  Unfortunately, the \hst\, composite
spectrum does not contain the \ion{Fe}{3} $\lambda 5129$ feature, so
we are unable to confirm the trend to higher redshifts and to
determine if the trend is the result of a selection effect.

  \subsection{Phase-Binned Spectra}\label{ss:age}

\begin{figure}
\epsscale{0.9}
\rotatebox{90}{
\plotone{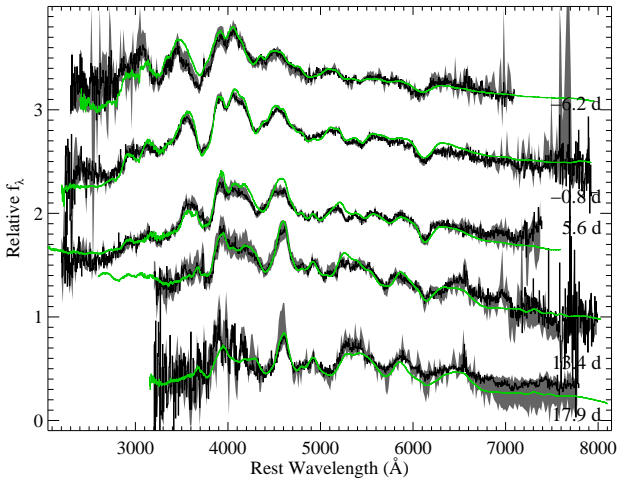}}
\caption{ESSENCE composite SN~Ia spectra for different phase bins.  The
spectra have phase bins of $-11 < t < 3$~d, $-3 < t < 3$~d, $3 < t <
10$~d, $10 < t < 17$~d, and $17 < t < 23$~d, with average phases of
$-6.2$~d, $-0.7$~d, 5.7~d, 13.3~d, and 17.9~d, respectively.  The $-3
< t < 3$~d spectrum is the same as in Figure~\ref{f:essmax_comp}.  The
grey regions are the $1\sigma$ boot-strap variation.  The green lines
are Lick composite spectra with the same phase bins and similar average
phases.}\label{f:age_comp}
\end{figure}

Another subsampling we created was based on phase, as estimated from
the light curves.  We present five phase-binned composite spectra in
Figure~\ref{f:age_comp}, corresponding to roughly one week before
maximum light, maximum light, and one, two, and three weeks after
maximum light, having phase bins of $-11 < t < -3$~d ($\mean{t} =
-6.2$~d), $-3 < t < 3$~d ($\mean{t} = -0.8$~d), $3 < t < 10$~d
($\mean{t} = 5.6$~d), $10 < t < 17$~d ($\mean{t} = 13.3$~d), and $17 <
t < 23$~d ($\mean{t} = 17.9$~d), respectively. Again, all ESSENCE
composites are generally similar to the Lick composites with some
differences.

The $-6.2$~d ESSENCE spectrum displays three noteworthy differences: a
higher \ion{Si}{2} $\lambda 6355$ velocity (12,800 \kms versus 12,100
\kms for the Lick spectrum), a slightly stronger \ion{Si}{2} $\lambda
6355$ line, and a UV excess.  The ESSENCE spectrum has a slightly
older average phase ($-6.2$~d compared to $-6.6$~d) and slightly
smaller average $\Delta$ ($-0.14$ compared to $-0.08$).  Although
these differences are small, they may account for part of the
different Si velocity and line strength.  For some SNe, our redshift
errors can be as large as \about 3000 \kms, which is much larger than
the difference between the two velocity measurements.  However, the other
lines in the spectrum, including the \ion{Si}{2} $\lambda 4130$ line,
do not show this velocity shift, indicating that this is intrinsic to
the \ion{Si}{2} $\lambda 6355$ line.  The UV excess is very difficult
to explain as merely a systematic error, although see
Section~\ref{ss:lines} for a discussion of the possible issues
regarding the treatment of the galaxy-light contamination.

The $-0.8$~d maximum-light ESSENCE spectrum is discussed in
Section~\ref{ss:maxlight}.

The 5.6~d ESSENCE spectrum is similar to the comparable Lick spectrum,
but there are more differences than in the earlier spectra.  The blue
wing of the \ion{S}{2} doublet ($\lambda 5612$) is stronger and the
overall continuum matching is worse than with other spectra.  This is
likely the result of poor galaxy-contamination correction.  Later-epoch 
spectra require starburst-galaxy spectra for the Lick spectra to
match the continua of the ESSENCE spectra.  The 5.6~d spectrum may be
the transition between these two regimes.

The 13.3~d and 17.9~d ESSENCE spectra are also similar to their Lick
counterparts.  However, to make this match, it was necessary to
include a percentage of starburst-galaxy spectra.  Restricting our
fitting method to normal galaxies, all features match between the
ESSENCE and Lick spectra, but their strengths are very different, with
the Lick features being much stronger.  With only normal galaxy light,
the ESSENCE spectra have very large blue excesses.  Unfortunately,
these bins contain lower-redshift SNe, and as such, we are unable to
probe the UV portion of the spectrum.

  \subsection{$\Delta$-Binned Spectra}\label{ss:delta}

To test the possibility that a certain luminosity-based subsample may
deviate from the local comparison, we separated the objects into three
luminosity bins defined by \citet{Jha06} of $\Delta < -0.15$, $-0.15 <
\Delta < 0.3$, and $\Delta > 0.3$, corresponding to overluminous,
normal, and underluminous objects, respectively.

\begin{figure}
\epsscale{0.9}
\rotatebox{90}{
\plotone{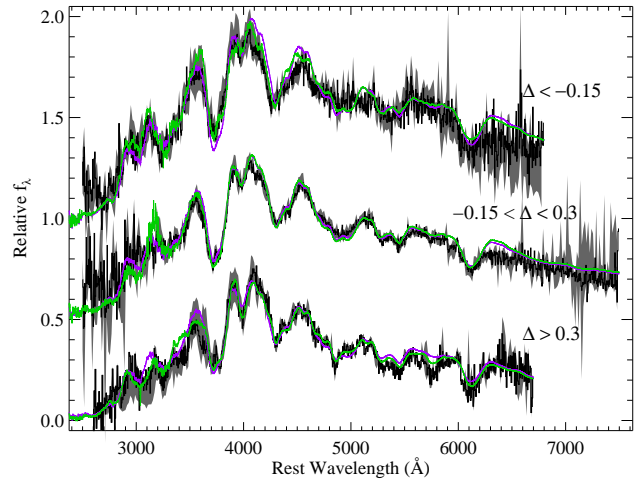}}
\caption{ESSENCE maximum-light composite SN~Ia spectra for different
$\Delta$ bins.  The composite spectra consist of 3 (10), 14 (18), and
15 (9) individual spectra with average $\Delta$ of 0.33, (0.43), 0.01
($-0.05$), and $-0.32$ ($-0.28$) for the underluminous, normal, and
overluminous subsamples defined by \citet{Jha06} ($\Delta > 0.3$,
$-0.15 < \Delta < 0.3$, and $\Delta < -0.15$) for the ESSENCE (Lick)
sample, respectively.  All have average redshifts of \about0.3.  The
grey regions are the $1\sigma$ boot-strap variation.  The green lines
are the Lick composite comparison spectra.}\label{f:delta_comp}
\end{figure}

In Figure~\ref{f:delta_comp}, we present the luminosity-binned
maximum-light ESSENCE composite spectra as well as the total and
luminosity-binned Lick spectra for comparison.  The ESSENCE and Lick
spectra are all generally consistent with both the total and
luminosity-binned Lick spectra.  The total and luminosity-binned
spectra are similar, with the total and normal-luminosity spectra
being nearly identical.  This indicates that the total composite
spectra of Lick (and to some degree the ESSENCE spectra) are a proxy
for the normal-luminosity SNe~Ia.

The normal-luminosity ESSENCE and Lick spectra are very similar,
with the spectra having minor differences.  The underluminous spectra
are also very similar; the main difference is that the \ion{Si}{2}
$\lambda 5972$ line is slightly stronger in the ESSENCE composite.
Since the underluminous composite is constructed from only three
spectra, this difference might not persist for a larger sample.
Although the overluminous spectra are generally similar, there are
minor differences between the Lick and ESSENCE composite spectra.  The
\ion{Fe}{2} 4800 feature, which includes the \ion{Fe}{3} $\lambda
5129$ line, is weaker in the ESSENCE spectrum, practically vanishing.

\section{Discussion}\label{s:discussion}

Section~\ref{s:results} shows that in every instance, the overall
appearance of the low-redshift Lick and high-redshift ESSENCE spectra
is very similar, with only a few small but notable differences.  In
multiple cases, the \ion{Fe}{2} absorption at \about3000~\AA\ present
in the Lick composite spectra is missing in the ESSENCE composite
spectra.  Similarly, the \ion{Fe}{2} 4800 feature is weaker in the
ESSENCE spectra, particularly at higher redshift and smaller $\Delta$
(that is, overluminous SNe~Ia).
The \ion{Si}{2} $\lambda 6355$ line is perhaps slightly stronger in
the premaximum ESSENCE composite spectrum, but weaker in others.  In
the premaximum ESSENCE composite spectrum, there is a UV excess and
the \ion{Si}{2} $\lambda 6355$ line is blueshifted more compared to
the Lick composite spectrum.  In this section, we attempt to explain
the physical nature as well as some consequences of these differences.

  \subsection{Evolution vs.\ Changing Demographics}

If there is a true spectroscopic difference between low and high-redshift
SNe~Ia, it is still necessary to determine if objects with the same 
value of $\Delta$ (and hence the same light-curve shape, to first order)
are changing with redshift (evolution) or if the population of the objects 
is changing with redshift (changing demographics). If, for instance, SNe 
with a given value of $\Delta$ are changing with redshift as a
result of metallicity differences, we would call that ``evolution.''  On
the other hand, if the metallicity is lower in star-forming galaxies and
more SNe~Ia occur in star-forming galaxies at high redshift, causing the
average SN to have lower metallicity, we would call that ``changing
demographics.''

A difference in samples caused by both changing demographics
and evolution are possibilities.  Photometric information suggests
that SN~Ia demographics do change with redshift \citep{Howell07};
namely there are more objects with higher stretch at high redshift.
Until we compare objects with similar conditions (host galaxies, delay
times, etc.), we will not be able to distinguish between these
possibilities.

  \subsection{Line Strengths}\label{ss:lines}

    \subsubsection{\ion{Fe}{2} 3000}

The \ion{Fe}{2} \about 3000~\AA\ absorption is absent in the
normal-luminosity ESSENCE composite spectrum, but the feature has
similar strength in the overluminous ESSENCE and Lick composite
spectra.  Although the \ion{Fe}{2} 4800 feature is also weaker at
higher redshift (because of a weaker \ion{Fe}{3} $\lambda 5129$ line,
which is a component of the \ion{Fe}{2} 4800 feature), the opposite is
true, with the line being weaker in the overluminous ESSENCE composite
spectrum but having similar strength in the normal luminosity ESSENCE
and Lick composite spectra.  It is therefore unlikely that the lack of
the \ion{Fe}{2} \about 3000~\AA\ absorption is the result of an
ionization or density effect (although there may be non-LTE effects).
Alternatively, there may be another species contributing to the
feature.

\begin{figure}
\epsscale{1.1}
\rotatebox{90}{
\plotone{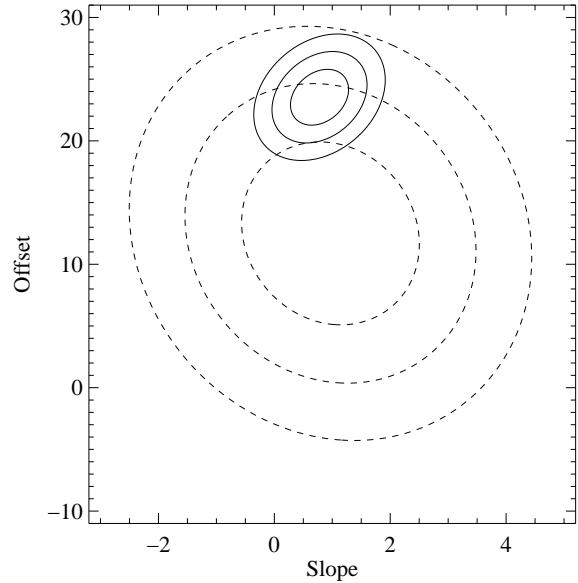}}
\caption{$1\sigma$, $2\sigma$, and $3\sigma$ confidence contours for the 
linear fit parameters for the \about 3000~\AA\ feature.  The smaller, solid
contours are for the Lick sample.  The larger, dashed contours are for
the ESSENCE sample.}\label{f:cont3000}
\end{figure}

To determine the significance of the differences in the strength of
this feature, we measured the line in individual spectra.  After
adding the appropriate amount of galaxy light (measured from the fit
to the spectra), we measured the pEW of all low-redshift spectra.  We
then fit a line to the pEWs as a function of age for $-11 < t < 11$~d.
We also performed a linear fit for the ESSENCE data over the same age
range.  In Figure~\ref{f:cont3000}, we show the confidence contours of
the fit parameters for the Lick and ESSENCE samples.  The contours
overlap, indicating that the differences in the samples are not
significant.

\begin{figure}
\epsscale{1.1}
\rotatebox{90}{
\plotone{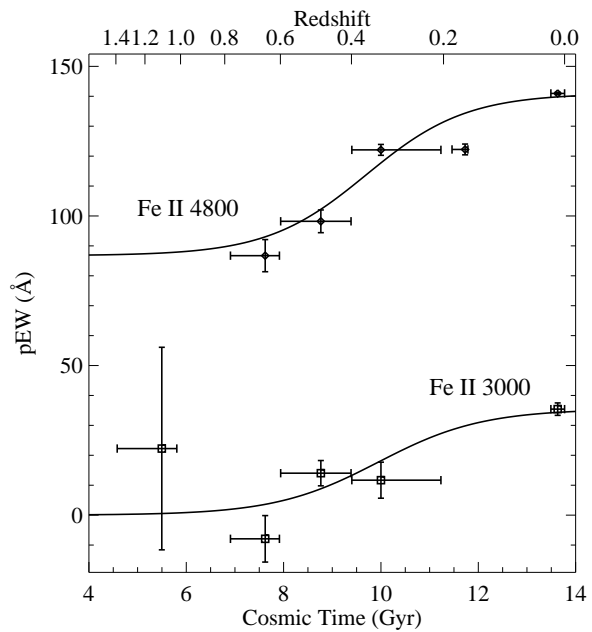}}
\caption{Psuedo-equivalent widths of the \about3000~\AA\ (squares) and
\ion{Fe}{2} 4800 (diamonds) features as a function of cosmic time.
The highest-redshift bins are consistent with $\text{pEW \!(3000~\AA)}
= 0$.  Overplotted are the best-fit \citet{Verhulst1845} curves
showing a transition from weaker to stronger features at $t_{tr, 3000}
= 9.9$ Gyr ($z_{tr, 3000} = 0.34$) and $t_{tr, 4800} = 9.7$ Gyr ($z_{tr,
4800} = 0.36$).  The pEW errors shown are only statistical; systematic
errors are typically 10~/AA.  The decrease in the \ion{Fe}{2} 4800
feature may be artificially smaller at higher redshift.  See
Figure~\ref{f:fepoints} for measurements of individual
spectra.}\label{f:pews}
\end{figure}

Nevertheless, we do see some indications of differences in the
strength of the line with redshift.  As seen in Figure~\ref{f:pews},
the Lick composite spectrum shows a strong line, while the $z > 0.6$
ESSENCE and \hst\, spectra show no indication of a line.  The
lower-redshift ESSENCE composite spectra also show weaker lines.  We
note that visually, there appears to be a slight absorption in the
$\mean{z} = 0.33$ spectrum, but nothing distinct in the $\mean{z} =
0.48$ spectrum.  There may be absorption in the \hst\, composite
spectrum (see Section~\ref{ss:redshift} and
Figure~\ref{f:redshift_comp}), but it is consistent with no
line.  Considering the systematic uncertainties, it is possible that
the data can be fit with a straight line and there is no change with
redshift.


%

In Figure~\ref{f:pews}, we plot the evolution of pEW of the \about
3000~\AA\ feature with cosmic time.  Using a \citet{Verhulst1845}
function, which is a logistic function originally used to describe
population growth, we fit the pEWs assuming $\text{pEW}(t = 0) = 0$
and $\text{pEW}(t = \infty) = \text{pEW}(z = 0)$, resulting in
\begin{equation}
  \text{pEW \!(3000 \AA)} = \frac{112 e^{0.938 t}}{3.54 \times 10^{4} +
    3.16 \left ( e^{0.938 t} - 1 \right )} \text{ \AA,}
\end{equation}
\noindent
which is also shown in Figure~\ref{f:pews}.  This equation yields a
transition time, $t_{tr} = 9.9$ Gyr, where the feature is at half its
current strength.  This corresponds to $z_{tr} = 0.34$.  However,
considering systematic errors, we believe that the feature is
consistent with having no change over time.

The analysis of the \about 3000~\AA\ absorption is hampered by a lack
of low-redshift UV spectra.  Examining individual spectra, we see a
range of strength from strong absorption
\citep[SN~1992A; ][]{Kirshner93} to emission at this wavelength
\citep[SN~1990N; ][]{Jeffery92}.  All low-redshift objects (including
those without light-curve information, which are not included in any
composite spectra) show this feature in absorption in at least one
spectrum.  However, because this wavelength can show both absorption
and emission, a composite spectrum from SNe~Ia could produce a lack of
absorption (similar to the ESSENCE spectra).  Without a better
understanding of this feature and its strength in the low-redshift
sample, we do not have the ability to make any claims of its variation
between low and high redshift.

    \subsubsection{\ion{Fe}{3} 5129}\label{sss:fe3}

\begin{figure}
\epsscale{1.1}
\rotatebox{90}{
\plotone{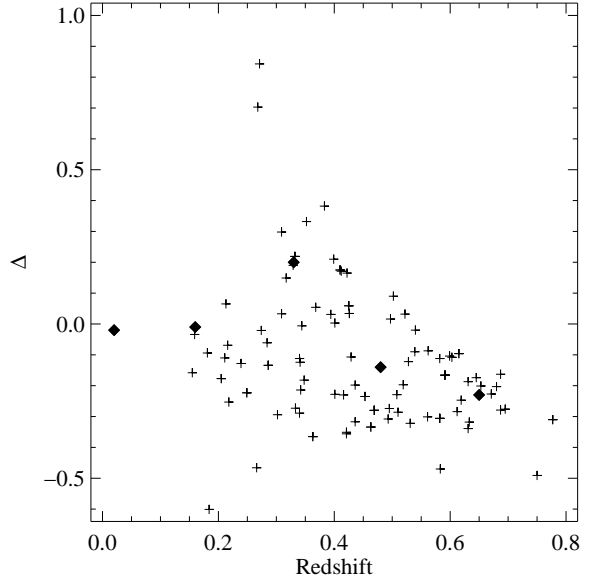}}
\caption{$\Delta$ distribution of ESSENCE objects as a function of
redshift.  The crosses are values for individual ESSENCE objects,
while the filled diamonds are the values for the redshift-binned
composite spectra (including the Lick composite
spectra).  The lack of high-$\Delta$ objects at high redshift is the
result of selection bias.}\label{f:zdelta}
\end{figure}

The one substantial difference between the low-$\Delta$ ESSENCE and
Lick composite spectra is the strength of the \ion{Fe}{2} 4800
feature.  The \ion{Fe}{2} 4800 feature is a blend of many lines, with
the strongest being \ion{Fe}{2}, \ion{Si}{2}, and \ion{Fe}{3}.
Examining Figures~\ref{f:essmax_comp} and \ref{f:redshift_comp}, we
see that the major difference in this feature between the ESSENCE and
Lick composite spectra is the reddest portion, attributed to
\ion{Fe}{3} $\lambda 5129$.  This feature is weaker in both the
higher-redshift and the low-$\Delta$ binned ESSENCE spectra.  The
average redshift for the low-$\Delta$ composite spectrum ($z = 0.51$)
is significantly larger than the total composite spectrum ($z = 0.37$)
and the normal-luminosity composite spectrum ($z = 0.27$).  Similarly,
as seen in Figure~\ref{f:zdelta}, the average $\Delta$ decreases with
redshift ($\Delta = -0.05$, 0.20, $-0.14$, and $-0.23$ for $0 < z <
0.2$, $0.2 < z < 0.4$, $0.4 < z < 0.6$, and $z > 0.6$, respectively).

  Note that the Lick and ESSENCE low-$\Delta$ composite spectra have 
the same average $\Delta$, $\mean{\Delta} = -0.27$.  Therefore, the
differences are not the effect of comparing different average
parameters. However, based on our current data set, we cannot
determine if (1) on average the \ion{Fe}{3} $\lambda 5129$ line is
weaker in all SNe~Ia with increasing redshift, or (2) the feature is
weaker in low-$\Delta$ SNe~Ia at high redshift and the increasing
influence of low-$\Delta$ objects on the sample for increasing
redshifts.  Because of this ambiguity, we cannot distinguish between
evolution and changing demographics.  To address this issue, we would
need a sufficiently large sample of $\Delta > -0.15$ SNe~Ia at high
redshift and $\Delta < -0.15$ objects over the entire redshift range.
This would allow us to create $\Delta-z$ binned composite spectra.
Unfortunately, our eleven low-$\Delta$ spectra and eight $\Delta >
-0.15$, $z > 0.4$ (two with $z > 0.6$) spectra do not allow this
binning.

Fitting the \citet{Verhulst1845} function to the \ion{Fe}{2} 4800
feature, assuming that $\text{pEW} (t = 0) = \text{pEW} (z = 0.65)$
and $\text{pEW} (t = \infty) = \text{pEW} (z = 0)$, we find
\begin{equation}
  \text{pEW \!(\ion{Fe}{2} 4800)} = \frac{226 e^{0.976 t}}{4.18 \times
    10^{4} + 5.42 \left ( e^{0.976 t} - 1 \right )} + 86.7 \text{ \AA.}
\end{equation}
\noindent
This equation has a transition time of $t_{tr} = 9.7$ Gyr,
corresponding to $z_{tr} = 0.36$.

As with all equivalent width measurements, one needs to examine if the
line strength is changing or if the continuum level is changing.  In
our case, there are two sources of continuum: the strength of the
``emission'' lines to either side of a feature, and the galaxy light.
In order to match the pEW(\ion{Fe}{2} 4800) values for the $z > 0.6$
ESSENCE and Lick composite spectra, the galaxy-light component of the
ESSENCE spectrum would need to be underestimated by 44\%.  It is
possible that the ``emission'' features on either side of the \ion{Fe}{2}
4800 feature are weaker in the ESSENCE spectra; however, the remarkably
similar maximum-light spectra at all points in the range 4000--6000~\AA\
(except for this feature), as seen in Figure~\ref{f:essmax_comp}, argues
against this.

\begin{figure}
\epsscale{0.9}
\rotatebox{90}{
\plotone{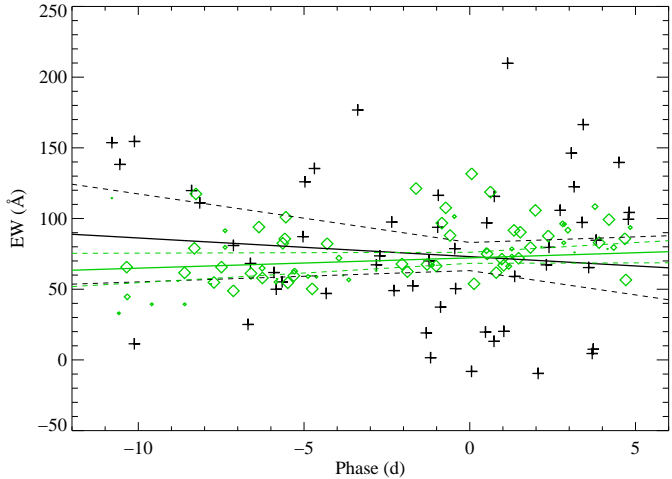}}
\caption{The EW measurements of \ion{Fe}{2} 4800 for indiviudal
spectra in the ESSENCE and Lick spectra.  The black crosses are for
ESSENCE spectra and the green diamonds are for Lick data.  The size of
the points is proportional to their weight.  The best-fit lines are
solid black and green lines for ESSENCE and Lick, respectively.  The
dashed lines are the 1$\sigma$ errors in the fits.  Each Lick
spectrum was given a weight of $1/N$, where $N$ is the number of
spectra from that object used in the fit.  This weights each object,
rather than each spectrum, equally.}\label{f:fepoints}
\end{figure}

\begin{figure}
\epsscale{1.1}
\rotatebox{90}{
\plotone{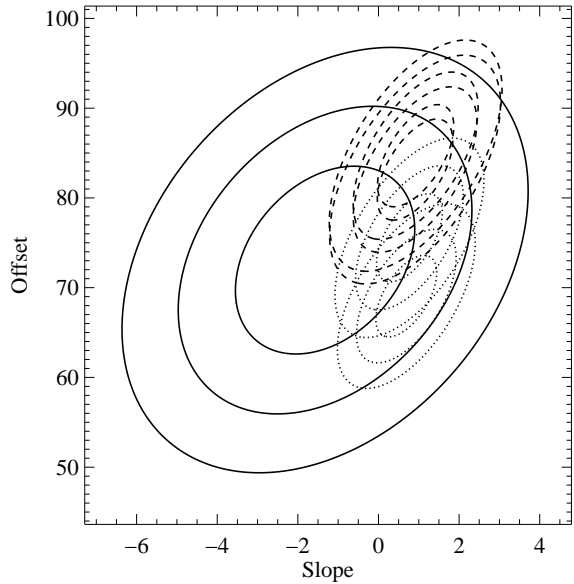}}
\caption{1$\sigma$, 2$\sigma$, and 3$\sigma$ confidence contours for the 
linear fit parameters for the \ion{Fe}{2} 4800 feature.  The large, solid
contours are for the ESSENCE sample and the dashed and dotted contours
are for the Lick sample.  The dashed and dotted contours are from the
1$\sigma$ extremes for galaxy contamination using the Sb and Sc
templates, respectively.  The contours from the best-fit galaxy
contamination are not plotted for clarity, but are close to averages
between the plotted contours.}\label{f:contfe}
\end{figure}

As we did with the \about 3000~\AA\ feature, we fit the pEW
measurements from individual spectra as a function of time for both
samples.  The EW measurements for the ESSENCE and Lick samples (with 
galaxy light added to best match the ESSENCE sample) are presented in
Figure~\ref{f:fepoints}.  We only look at the data for $t < 5$~d,
which is both where the data are linear and where the line is dominated
by the same absorption features.  Linear fits were performed for both the
Lick and ESSENCE samples with the best fits shown in
Figure~\ref{f:fepoints}.  The confidence contours associated with the
linear fits are plotted in Figure~\ref{f:contfe}.  The two samples are
different at less than the 1$\sigma$ level.  We also considered the
Lick sample using the PCA-reconstructed galaxy spectrum instead of the
template-galaxy spectrum.  As seen in Figure~\ref{f:essmax_pca}, this
results in a slightly smaller pEW.  Regardless of the galaxy template
used, with any reasonable value of galaxy contamination the
samples are still different at the \about 2$\sigma$ level.

The galaxy contamination is a difficult issue.  Ideally, we would be
able to remove the galaxy contamination in each spectrum.  However,
with the low S/N and small wavelength range of each individual spectrum
and the uncertainty in the amount of galaxy contamination, we would need
to make large assumptions about the overall continuum shape of each SN
spectrum.  Our method of matching the composite spectra
avoids these issues by comparing high S/N spectra covering a large
wavelength range.  A test of the accuracy of this method is to examine
the photometric fluxes of the galaxies and SNe at maximum light.  From
our light curves, we are able to determine the galaxy contamination in
the observed $R$ band.  Then, assuming the SN spectrum is the Lick
maximum-light composite spectrum, we can determine the galaxy
contamination in the \ion{Fe}{2} 4800 feature.  Doing this, we find
that the galaxy contamination has median and mean values of 17--19\%
and 22--23\%, respectively, regardless of galaxy template.  The
1$\sigma$ upper limit is 47--60\% for all galaxy templates.
Considering that we remove some galaxy light from our spectra while
reducing our data, we regard these values as upper limits.  This
is consistent with what we have found from matching the low and
high-redshift composite spectra.

\begin{figure}
\epsscale{1.1}
\rotatebox{90}{
\plotone{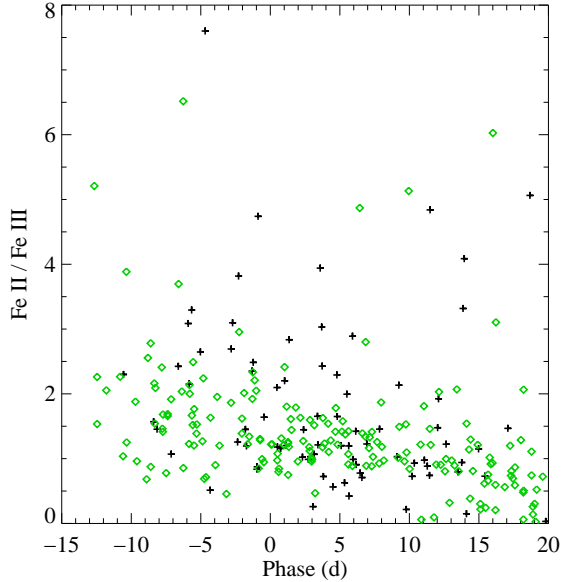}}
\caption{The \ion{Fe}{2}/\ion{Fe}{3} ratio for spectra in the ESSENCE
(black crosses) and Lick (green diamonds) samples.}\label{f:feratio}
\end{figure}

A separate way of examining the Fe lines which avoids the majority of
galaxy contamination is a ratio of the depths of the \ion{Fe}{2} and
\ion{Fe}{3} lines.  We measured the depths of these lines from a line
segment connecting the maxima on either side of the feature, similar to 
the method of determining the $\mathcal{R}$(Si) ratio first presented by
\citet{Nugent95}.  Since the ESSENCE spectra have low S/N, finding a
true minimum for the \ion{Fe}{3} line is difficult.  We therefore
measured the depth of \ion{Fe}{3} for all spectra at the wavelength of
the minimum in the Lick maximum-light composite spectrum.  If this
wavelength is slightly off the true minimum, causing the flux of the
edge of the line to be measured, we would underestimate the true
\ion{Fe}{3} depth, leading to a larger \ion{Fe}{2}/\ion{Fe}{3} ratio
for any given spectrum; however, since all spectra were treated in the
same way, this should not significantly influence the results.  The values 
of this ratio are presented in Figure~\ref{f:feratio}, showing that
this ratio evolves linearly with age, with the ESSENCE sample having a
slightly larger \ion{Fe}{2}/\ion{Fe}{3} ratio on average.  However, if
we perform a two-dimensional Kolmogorov-Smirnov test, the samples are
different at the 95\% level, consistent with a \about2$\sigma$
difference.

There are two possible physical explanations for a weaker \ion{Fe}{3}
line in high-redshift SNe~Ia.  The first is that high-redshift SNe~Ia
have lower photospheric temperatures (on average) than their low-redshift
counterparts. The typical photospheric temperature of maximum-light
SNe~Ia is 10,000~K, which is very close to the transition between
\ion{Fe}{2} dominance and \ion{Fe}{3} dominance \citep{Hatano99}.  A
temperature change of 1000~K can dramatically change the ratio of the
Fe lines.  This effect has been used to explain the strong \ion{Fe}{3}
lines in SN~1991T \citep{Filippenko92:91T}.

An alternative explanation is that the weaker \ion{Fe}{3} line is the
result of lower metallicity.  \citet{Hoflich98} shows that the
physical quantity most sensitive to changing metallicity is the
$^{54}$Fe production.  As metallicity decreases, $Y_{e}$, the electron
fraction, increases.  This, in turn, causes less $^{54}$Fe production.
\citet{Hatano02} suggests that the distribution of the Fe isotopes
within the SN ejecta cause \ion{Fe}{2} lines to be generated by
$^{56}$Fe, while \ion{Fe}{3} lines come from $^{54}$Fe.  Therefore,
the weaker \ion{Fe}{3} $\lambda 5129$ line may be an indication of
lower metallicity for high-redshift SN~Ia progenitors.
\citet{Sauer08} offers a slightly different explanation: as
metallicity decreases, there is less backwarming from UV photons,
which decreases the temperature and significantly reduces the
\ion{Fe}{3}/\ion{Fe}{2} ratio.  In both models, lower metallicity has
the same effect: weaker \ion{Fe}{3} lines.

Finally, the red portion of the entire feature may be dominated by a
species different from \ion{Fe}{3}, such as \ion{Si}{2} or
\ion{Fe}{2}.  Although this is a possibility, it would require small
blueshifts ($v < 5000$ \kms) for those lines.

    \subsubsection{\ion{Si}{2} 6355}

A visual inspection of the \ion{Si}{2} $\lambda 6355$ line shows that
the placement of the continuum of the red wing of
the line is causing the differences between the ESSENCE and Lick
maximum-light composite spectra.  Although this may be a real
difference, the galaxy light is comparable to the SN component
for wavelengths redward of 6000~\AA.  Therefore, the comparison of the
ESSENCE and Lick spectra is highly dependent on the galaxy spectrum.
If we examine the feature when using the PCA-reconstructed galaxy
spectrum to contaminate the Lick composite spectrum, the difference is
much smaller.  Considering this, and also that other lines from
intermediate-mass elements (including \ion{Si}{2} $\lambda 4130$) do
not show different pEWs between the Lick and ESSENCE spectra, we do
not believe the \ion{Si}{2} difference is significant.  We do,
however, suggest that the differences in the premaximum \ion{Si}{2}
$\lambda 6355$ line (outlined in Section~\ref{ss:metallicity}) are
probably real and significant.

%
%
%
%
%
%
%

  \subsection{Metallicity}\label{ss:metallicity}

The two best observables for determining metallicity differences in
SN~Ia spectra are the velocity and depth of the \ion{Si}{2} $\lambda
6355$ line and the UV flux level \citep{Hoflich98, Lentz00}.  As
metallicity increases, the opacity of the ejecta increases, and 
consequently the velocity of the \ion{Si}{2} line becomes more 
blueshifted while the line strengthens.  As already discussed in
Section~\ref{s:intro}, the effect of changing metallicity on the UV
continuum is disputed.  Rather than trying to determine the correct
model, we first note that all models show a change in the UV continuum
with changing metallicity.  In the \citet{Lentz00} model, with
increasing metallicity, there is more metal-line cooling, creating a
redder SED, and at the same time, there is more line blanketing from
metal lines in the blue, absorbing the flux at those wavelengths.  
The combination of these effects creates a UV deficit at high
metallicity. Moreover, differing density structures can cause these
lines to form at smaller radii, decreasing the UV flux level with
increasing metallicity \citep{Hoflich98,Dominguez01}.

As described by \citet{Lentz00}, most metallicity effects are easier
to observe at early times.  As seen in Section~\ref{ss:age}, the
premaximum ESSENCE spectrum has both a stronger \ion{Si}{2} $\lambda
6355$ line with a larger velocity, and a UV excess, compared to the Lick
premaximum composite spectrum.  Unfortunately, the wavelength region
where we can easily differentiate the models by UV flux is at \about
2000~\AA, which is blueward of our spectra.  However, the general
trend of the \citet{Lentz00} models is that the UV flux level
decreases with increasing metallicity.  If this is the case, then the
ESSENCE objects would have lower metallicity than the Lick objects.
However, using the results of \citet{Hoflich98} one would arrive at the
opposite conclusion --- namely, the ESSENCE objects have higher
metallicity than the Lick objects.  According to the \citet{Lentz00}
models, the stronger, more-blueshifted \ion{Si}{2} $\lambda 6355$
in the ESSENCE composite spectrum is suggestive of higher
metallicities, conflicting with the UV excess.

The idea of higher metallicity for high-redshift SNe~Ia is
counterintuitive.  However, there is a possible explanation if we
consider the two-channel model for SN~Ia progenitors
\citep{Mannucci05, Scannapieco05}.  Examining an extreme example,
suppose that at $z = 0$ all SNe~Ia come from a delayed channel and at
$z = 0.5$ all SNe~Ia come from a prompt channel \citep[an
approximation of reality;][]{Sullivan06}, which we will assume to have
no delay time for simplicity.  Since the light travel time from $z =
0.5$ to $z = 0$ is 5 Gyr, if the delay time for the delayed channel is
$>5$ Gyr, then the low-redshift SNe~Ia would come from 
lower-metallicity progenitors. However, the data suggest that the 
difference in delay times is \about 2 Gyr \citep{Scannapieco05}.

There are two major possible sources of systematic error.  First, the
shape of the UV continuum is, to some degree, dependent on the galaxy 
spectrum we fit to the ESSENCE spectrum.  If we have used the incorrect 
galaxy type or an incorrect amount of galaxy light (both of which may vary
over our wavelength range), our UV continuum could be undersubtracted,
causing the UV excess.  However, the UV flux of galaxies is small
compared to the optical flux. Hence, if we have undersubtracted galaxy
light in the UV, we would expect the effect to be even stronger in the
optical, which it is not.  However, if our lower-redshift premaximum
spectra tended to come from SNe in early-type galaxies and our
higher-redshift premaximum spectra tended to come from SNe in
late-type galaxies, then the UV portion of the spectrum (which
is only visible in the higher-redshift spectra) could be contaminated
by relatively bluer galaxy light, for which we may not correctly
account.

The other possible source of systematic error is the scarcity of
premaximum low-redshift UV spectra.  There are 6 spectra contributing
to the UV portion of the Lick premaximum composite spectrum (1 from
SN~1989B, 4 from SN~1990N, and 1 from SN~1992A).  If these objects are
atypical, having less UV flux than most SNe~Ia at early times, the
UV ``excess'' in the ESSENCE spectrum would simply be an artifact of
comparing to these atypical objects.  An argument against this is that
spectra of SNe~1990N and 1992A are included in other Lick composite
spectra with different phase binning (such as the maximum-light
spectrum), and they do not have significantly depressed UV flux.

This result requires us to look at the other spectra in detail,
searching for additional clues of a difference in metallicity.  As seen 
in Figures~\ref{f:essmax_comp}, \ref{f:redshift_comp}, \ref{f:age_comp},
and \ref{f:delta_comp}, there are no other obvious UV excesses or
extraordinary line velocities.  As discussed in Section~\ref{sss:fe3},
the difference in the \ion{Fe}{3} line may be the result of differing
metallicity.  Since the outer layers of the progenitor are most
affected by metallicity and these layers are only seen at very early
times, a noticeable change of metallicity may not be possible for
times past maximum brightness. \citet{Dominguez01} suggests that the
$B - V$ color at maximum light decreases with increasing metallicity.
They show that $\Delta(B - V) = 0.05$ mag for metallicity increasing
from $\log(Z/Z_{\sun}) = -10$ to $-1.7$.  We measure $B - V = 0.22$
mag and 0.24 mag for the ESSENCE and galaxy-contaminated Lick
maximum-light composite spectra, respectively, suggesting that this
effect may be present in our spectra.  Unfortunately, the galaxy-light
contamination reduces our ability to measure accurate colors from the
spectra, so we cannot determine with any certainty if this effect is
real.

One place one might expect to see more pronounced evolutionary effects
is in the low-$\Delta$ objects.  Overluminous SNe~Ia are found in
star-forming galaxies \citep{Hamuy00, Howell01}.  Based on SN rates,
\citet{Mannucci05} and \citet{Scannapieco05} have suggested that there
exist two channels for SNe~Ia: short (\about 1 Gyr) and long ($\gtrsim
3$ Gyr) delays.  The SN rate from the short-delay channel is
proportional to star formation.  These data suggest that overluminous
SNe~Ia tend to come from a short-delay progenitor channel, from which
the progenitor is more biased by galactic environment, and thus
galactic evolution, than the long-delay channel.  As star-forming
galaxies increase their metallicities with time, they imprint that
information in the SN~Ia progenitors, creating somewhat different
spectra at high and low redshift.

Comparing the UV spectra of both the ESSENCE and Lick low-$\Delta$
composite spectra, we see no UV flux differences in a subsample that
should be particularly sensitive to these changes.  However, the
low-$\Delta$ sample is where the \ion{Fe}{3} $\lambda 5129$ line
changes the most between low and high redshift.  Although we cannot
say definitively if this is the result of evolution or changing sample
demographics, we see that the high-$\Delta$ Lick and ESSENCE composite
spectra have weak \ion{Fe}{3}, with the ESSENCE composite spectra
weaker than the Lick spectra.  If the high-$\Delta$ objects come from 
an old progenitor population, the ESSENCE low-$\Delta$ objects from an
intermediate-age population, and the Lick low-$\Delta$ objects from a
young population, the \ion{Fe}{3} line may be the tracer of the
metallicity evolution through these populations.  A more appropriate
subsampling, which might yield more definitive results, may be based on
the host-galaxy star-formation rate \citep{Sullivan06, Howell07}.




  \subsection{K-Corrections}

One of the largest systematic errors associated with SN distance
determinations is the uncertainty in the K-corrections.  Ideally,
one would have a time sequence of perfectly flux-calibrated,
contamination-free spectra (either for the object or for all template
objects) to synthesize light curves that perfectly match the observed
bands in the frame of the object.  Since getting multiple-epoch
spectra of numerous low-redshift objects or obtaining
multiple-epoch spectra of a single high-redshift SN is very time
consuming, we implement K-corrections to match photometric
observations of high-redshift SNe to those of low-redshift SNe.  If
there is a small difference between the SED of the high-redshift SN
and its low-redshift counterpart near the edge of the filter
transmission function, errors in the magnitudes (and thus overall
distance) are introduced.

One method of determining the K-corrections for a particular object
consists of using a template low-redshift spectrum, warping it to the
observed colors of the high-redshift SN, and then determining the
comparable rest-frame magnitudes \citep{Kim96, Nugent02}.  This method
is particularly sensitive to differences between SEDs.


In order to determine the impact of using one particular spectral
template for K-corrections, we warped the Nugent and SNLS templates
and the Lick and ESSENCE composite spectra to have the same rest-frame
colors, redshifted the spectra, and extracted synthetic photometry.
The warping is dependent on where the anchor points are placed, but
this is not critical for our applications.  We extracted $R$-band and
$I$-band magnitudes for all three spectra over a redshift range of $0
< z < 1$.  Because of its bluer rest-frame wavelength range, the
ESSENCE spectrum cannot be used for the lowest redshifts.
Furthermore, as seen in Figure~\ref{f:essmax_comp}, the large
uncertainty in the ESSENCE spectrum at red wavelengths makes the
$I$-band magnitudes for $z < 0.5$ unreliable.

\begin{figure}
\epsscale{0.9}
\rotatebox{90}{
\plotone{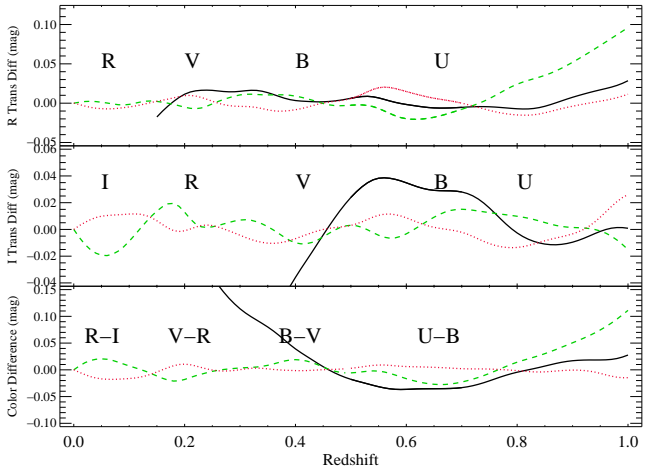}}
\caption{The difference between maximum-light synthetic rest-frame
\ubvri and K-corrected synthetic observed-frame $R$-band and $I$-band
photometry.  The black curve is the difference between the ESSENCE
composite spectrum and the Lick template.  The green (dashed) and red
(dotted) curves are the difference between the Lick composite spectrum
and the Nugent and SNLS templates, respectively.}\label{f:kcor}
\end{figure}

As seen in Figure~\ref{f:kcor}, all four spectra are generally
consistent to $< \pm 0.02$ mag.  The main deviation between the Lick
composite and Nugent template is at $z > 0.8$, where the near-UV is
redshifted into the $R$ band.  The differences between the spectra in
the UV (as seen in Figure~\ref{f:lick_nug}) create this large
difference.  The SNLS template is more consistent with both the Lick
and ESSENCE composite spectra.

This analysis shows that the systematic error associated with
K-corrections is small but non-negligible, with $\sigma_{K} \approx
0.01$--0.02 mag for $z < 1$, consistent with that found by
\citet{Hsiao07}, but larger than that assumed by \citet{Wood-Vasey07}.
We also note that the Nugent template has large differences compared
to both the Lick and ESSENCE composite spectra; it should not be used
for SNe with $z > 0.8$.

%
%
%
%

  \subsection{Constraining Evolution}

Ultimately, studies of SN evolution are aimed at determining how the
luminosity of the average SN~Ia changes with redshift.  The change in
luminosity will directly affect our measurements of SN distances, and
thus measurements of cosmological parameters.  Currently, the
systematic error associated with SN evolution is estimated to be $<
0.02$ mag, corresponding to an error of 0.02 in measuring $w$, the
equation-of-state parameter for dark energy.  With our composite
spectra, we have the first way to quantify this error.

\begin{figure}
\epsscale{0.9}
\rotatebox{90}{
\plotone{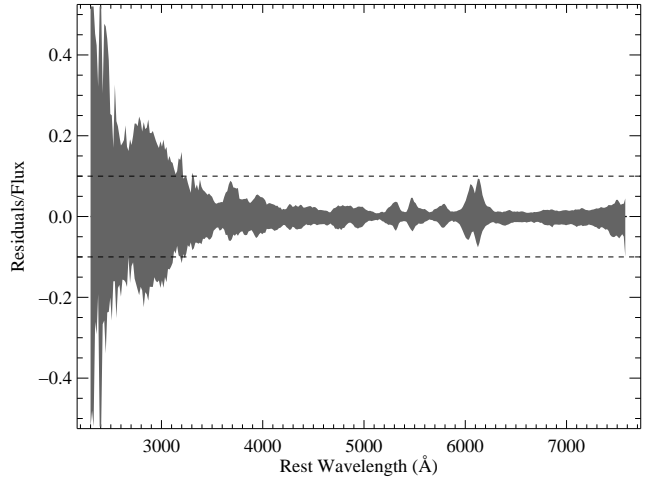}}
\caption{The percent variation in the Lick maximum-light composite
spectrum.  The dashed lines are the $\pm 10$\% variation.  The
variation is larger at shorter wavelengths.}\label{f:lickvar}
\end{figure}

In Figure~\ref{f:lickvar}, we show the percent variation (the ratio of
the 1$\sigma$ error residuals to the SN flux) for the Lick
maximum-light composite spectrum.  For most optical wavelengths, these
values are dominated by the intrinsic variation of the SN spectra.  In
the UV, however, the noise in the composite spectrum from the small
number of spectra dominates over the variance amongst objects.  It is
noteworthy that for wavelengths in the range of 3200--4000~\AA, the
variation is larger than for longer wavelengths and increasing with
smaller wavelengths.  This has been seen photometrically as a
relatively large variation in the $U$-band light curves of nearby
objects \citep{Jha06}.  We therefore expect this intrinsic variation
to continue to increase into the UV.

\begin{figure}
\epsscale{0.8}
\rotatebox{90}{
\plotone{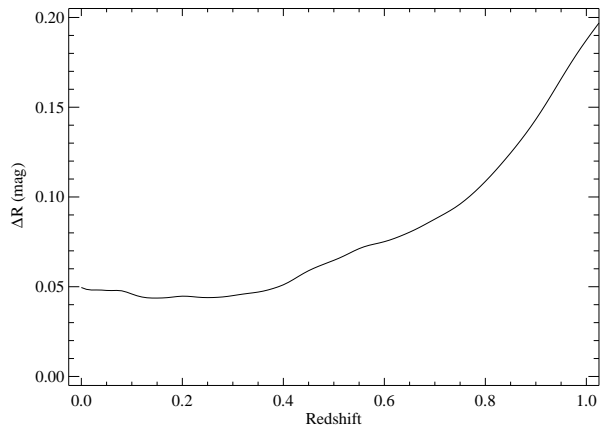}}
\caption{The difference in measured $R$-band magnitudes for
the upper and lower 1$\sigma$ variance Lick maximum-light composite
spectrum with redshift.  This difference indicates an upper limit on
the $R$-band difference after already normalizing the spectra as
described in Section~\ref{s:method}.}\label{f:revolution}
\end{figure}

We see that the intrinsic variation of SN spectra is at \about3\% for
most of the optical range.  Binning over larger wavelengths, the
variation in individiual features will average out, causing the extremes
of this variation to decrease slightly.  In Figure~\ref{f:revolution},
we show the difference in observed $R$-band magnitudes of the
1$\sigma$ variance spectra.  This measurement is an indication of the
variation of the spectra at maximum light and not the peak luminosity,
since we have normalized our spectra to have the same flux at a
particular wavelength.  For $z > 0.4$, we see that the difference
increases with redshift as the rest-frame UV is redshifted into the
observed-frame $R$ band.  This is probably an overestimate of the
maximum difference, since most spectra will vary within the 1$\sigma$
range rather than being at either the high or low end.

\begin{figure}
\epsscale{0.9}
\rotatebox{90}{
\plotone{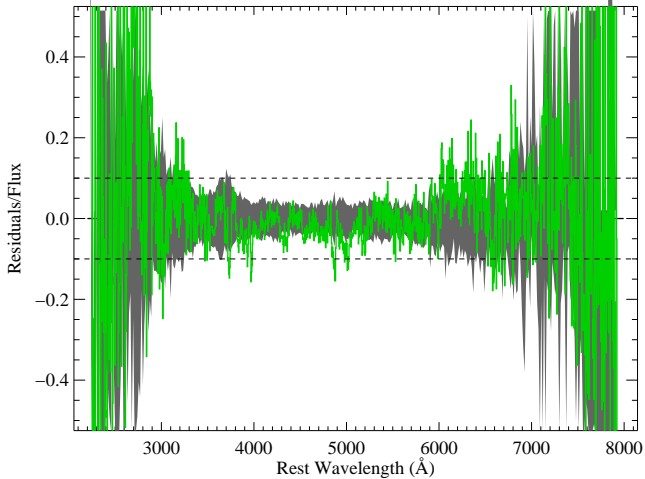}}
\caption{The percent variation in the ESSENCE maximum-light composite
spectrum in grey.  The green line is the percent variation of the
residuals of the Lick maximum-light composite spectrum (as seen in
Figure~\ref{f:essmax_comp}).  The dashed lines are the $\pm 10$\%
variation.}\label{f:essvar}
\end{figure}

We also present the percent variation for the ESSENCE maximum-light
composite spectrum in Figure~\ref{f:essvar}.  For most of the
rest-frame optical, the variation is \about5\%.  It is possible that
the high-redshift sample has a larger intrinsic variation than the
Lick sample, but it is more likely that the variation is dominated by
the different amounts of galaxy contamination and noise in the
spectra.  Figure~\ref{f:essvar} also shows that the difference between
the Lick and ESSENCE maximum-light composite spectra differs by
$\lesssim 10$\% over most of the rest-frame optical.  We can therefore
constrain evolution between our two samples to \about 10\% across the
spectrum.

\section{Conclusions}\label{s:conclusions}

By combining many low-S/N, high-redshift SN~Ia spectra, we are able to
construct the first series of composite SN~Ia spectra based on the
parameters of redshift, phase, $\Delta$, and $A_{V}$.  In addition, we
constructed similar composite spectra from a high-quality sample of
low-redshift SN~Ia spectra obtained over the last two decades.
Comparison of the composite spectra has shown that once we account for
galaxy-light contamination, the two samples are remarkably similar.
There are several minor deviations between low and high-redshift
samples.  These deviations fall into three categories:  related to
metallicity, related to $^{56}$Ni production, and unknown.

The UV excess in the ESSENCE premaximum spectrum is indicative of a
different metallicity for the low and high-redshift SNe.  Depending on
the model, a UV excess is the result of higher or lower metallicity
\citep{Hoflich98, Lentz00}.  The stronger, more blueshifted
\ion{Si}{2} $\lambda 6355$ line in the ESSENCE premaximum spectrum
indicates a higher metallicity \citep{Lentz00}.

The most significant difference between our samples is the varying
strength of the \ion{Fe}{3} $\lambda 5129$ line.  The evolution of the
\ion{Fe}{3} line may indicate that high-redshift SNe~Ia have lower
temperatures than low-redshift SNe~Ia.  This, in turn, suggests that
SNe~Ia should be less luminous at high redshift.  The weak \ion{Fe}{3}
line may indicate lower $^{54}$Fe production, which could be the
result of lower metallicity.  Alternatively, lower metallicity may
cause less backwarming from UV photons, decreasing the temperature and
the \ion{Fe}{3}/\ion{Fe}{2} ratio.  It is unclear if this difference
is from evolution in all SNe~Ia, evolution just in the low-$\Delta$
SNe~Ia, changing demographics, or a selection effect. Low-$\Delta$
SNe~Ia tend to come from the short-delay progenitor channel.  Because
of the short delay, the progenitors of these SNe are more biased by
galactic environment, and thus galactic evolution, than the
progenitors of long-delay channel SNe~Ia.  It is therefore not
surprising that the difference in the \ion{Fe}{3} line is more obvious
in the low-$\Delta$ objects.


It is possible that the different strengths of the \about3000~\AA\
\ion{Fe}{2} feature between low and high redshift is an artifact of
the construction of the composite spectra.   An analysis of the
individual spectra of both samples indicates that the samples are not
significantly different; however, the small number of UV objects
hampers this study.

It is difficult to definitively detect metallicity differences between
the Lick and ESSENCE samples.  First, we have observed three
differences between the samples which harbinger a difference in
metallicity: a UV excess, a stronger and more blueshifted \ion{Si}{2}
$\lambda 6355$ line, and a weaker \ion{Fe}{3} $\lambda 5129$ line.
The UV excess is an ambiguous indicator since the models disagree if
it indicates lower or higher metallicity.  The \ion{Si}{2} line in the
premaximum spectrum suggests higher metallicity, and the \ion{Fe}{3}
line suggests lower metallicity.

We have also shown that the previously published low-redshift template
spectra have multiple drawbacks when comparing to high-redshift
composite spectra.  We therefore caution against using these templates for
studies of SN~Ia evolution.  Furthermore, deriving K-corrections from
any low-redshift template is difficult; however, the systematic errors
are likely to be relatively small.

We see that the intrinsic variation of low-redshift SN spectra is
\about3\% in the optical.  The spectra vary more in the near-UV and UV
as suggested by photometry \citep{Jha06}.  We are able to put the
first constraints of SN~Ia evolution to $\lesssim 10$\%.

The results of this study are very suggestive, but require further
investigation.  In order to improve our understanding of SN~Ia
evolution, we propose three future studies related to this work.
First, the theoretical models of the effects of metallicity on SN~Ia
spectra should be expanded.  With the current ambiguity amongst
models, we cannot determine the direction of the trend in
metallicity.  Second, we should gather many more high-redshift spectra
to disentangle the redshift-$\Delta$ ambiguity.  Finally, further UV
observations of nearby SNe~Ia are desperately needed.  The only
current instrument available for the task is the {\it Swift} \uvot.
However, previous attempts at obtaining SN~Ia UV spectra have been
disappointing \citep{Brown05}.  We suggest an intense campaign
spending several hours per spectrum (similar to \iue) with the \uvot.
In the near future, we may once again have the ability to obtain
high-quality UV spectra with \hst\, using \stis\ or COS. If the upcoming
\hst\, servicing mission is successful, we strongly suggest a massive
campaign to observe local SNe~Ia in the UV.  Since \jwst\, does not have
the capabilities to observe the UV, this may be our last opportunity
for many years.

\begin{acknowledgments} 

Based in part on observations obtained at the Cerro Tololo
Inter-American Observatory, which is operated by the Association of
Universities for Research in Astronomy, Inc. (AURA) under cooperative
agreement with the National Science Foundation (NSF); the European
Southern Observatory, Chile (ESO Programmes 170.A-0519 and
176.A-0319); the Gemini Observatory, which is operated by the
Association of Universities for Research in Astronomy, Inc., under a
cooperative agreement with the NSF on behalf of the Gemini
partnership: the NSF (United States), the Particle Physics and
Astronomy Research Council (United Kingdom), the National Research
Council (Canada), CONICYT (Chile), the Australian Research Council
(Australia), CNPq (Brazil), and CONICET (Argentina) (Programs
GN-2002B-Q-14, GS-2003B-Q-11, GN-2003B-Q-14, GS-2004B-Q-4,
GN-2004B-Q-6, GS-2005B-Q-31, GN-2005B-Q-35); the Magellan Telescopes
at Las Campanas Observatory; the MMT Observatory, a joint facility of
the Smithsonian Institution and the University of Arizona; and the
F.~L. Whipple Observatory, which is operated by the Smithsonian
Astrophysical Observatory. Some of the data presented herein were
obtained at the W.~M. Keck Observatory, which is operated as a
scientific partnership among the California Institute of Technology,
the University of California, and the National Aeronautics and Space
Administration; the Observatory was made possible by the generous
financial support of the W.~M. Keck Foundation.  Most of the
low-redshift SN Ia spectra used here were obtained by A.V.F.'s group
over the course of two decades with the 3-m Shane reflector at Lick
Observatory. We thank the Lick staff for their dedicated help, as well
as many graduate students for assistance with the observations and
reductions.

The ESSENCE survey team is very grateful to the scientific and
technical staff at the observatories we have been privileged to use.

{\it Facilities:} 
  \facility{Blanco (MOSAIC II)}, 
  \facility{CTIO:0.9m (CFCCD)}, 
  \facility{Gemini:South (GMOS)}, 
  \facility{Gemini:North (GMOS)}, 
  \facility{Keck:I (LRIS)},
  \facility{Keck:II (DEIMOS, ESI)},
  \facility{Lick:3m Shane (UV Schmidt, Kast)},
  \facility{VLT (FORS1)},
  \facility{Magellan:Baade (IMACS)}, 
  \facility{Magellan:Clay (LDSS2)}.

The survey is supported by the US National Science Foundation through
grants AST--0443378 and AST--0507475. The Dark Cosmology
Centre is funded by the Danish National Research Foundation.  S.J.
thanks the Stanford Linear Accelerator Center for support via a
Panofsky Fellowship.  A.R. thanks the NOAO Goldberg fellowship program
for its support.  P.M.G. is supported in part by NASA Long-Term
Astrophysics Grant NAG5-9364 and NASA/HST Grant GO-09860.  R.P.K. enjoys
support from AST06-06772 and PHY99-07949 to the Kavli Institute for
Theoretical Physics.  A.C. acknowledges the support of CONICYT, Chile,
under grants FONDECYT 1051061 and FONDAP Center for Astrophysics
15010003. A.V.F.'s supernova group at U.C. Berkeley has been supported
by many NSF grants over the past two decades, most recently 
AST--0307894 and AST--0607485.

Our project was made possible by the survey program administered by
NOAO, and builds upon the data-reduction pipeline developed by the
SuperMacho collaboration.

\end{acknowledgments}

\bibliographystyle{apj}
\bibliography{astro_refs}

\clearpage
\LongTables

\begin{deluxetable}{lllrrlc}
\tablewidth{0pt}
\tablecaption{High-$z$ SN~Ia Information\label{t:hiz}}
\tablehead{
\colhead{ESSENCE} &
\colhead{IAU} &
\colhead{$z$} &
\colhead{Phase\tablenotemark{a}} &
\colhead{$\Delta$\tablenotemark{b}} &
\colhead{$A_{V}$} &
\colhead{Rest Wavelength} \\
\colhead{Name} &
\colhead{Name} &
\colhead{} &
\colhead{(d)} &
\colhead{} &
\colhead{(mag)} &
\colhead{Range (\AA)}}

\startdata

b010 & 2002iy  & 0.591 &  -5.0 & -0.166 & 0.104 & $2517 - 6461$ \\
b010 & 2002iy  & 0.591 &  -1.8 & -0.166 & 0.104 & $2130 - 5832$ \\
b010 & 2002iy  & 0.591 &  13.9 & -0.166 & 0.104 & $3469 - 5933$ \\
b010 & 2002iy  & 0.591 &  14.4 & -0.166 & 0.104 & $2756 - 5605$ \\
b013 & 2002iz  & 0.426 &  -1.0 &  0.034 & 0.170 & $2842 - 7190$ \\
b013 & 2002iz  & 0.426 &  19.8 &  0.034 & 0.170 & $3085 - 6255$ \\
b016 & 2002ja  & 0.329 &   0.1 &  0.190 & 0.359 & $2997 - 7725$ \\
b020 & 2002jr  & 0.425 &  -8.4 &  0.059 & 0.202 & $2810 - 7204$ \\
b020 & 2002jr  & 0.425 &  12.6 &  0.059 & 0.202 & $3684 - 6610$ \\
d033 & 2003jo  & 0.531 &   4.8 & -0.322 & 0.085 & $2779 - 5833$ \\
d058 & 2003jj  & 0.583 &   1.1 & -0.470 & 0.119 & $2736 - 5641$ \\
d083 & 2003jn  & 0.333 &   3.8 & -0.273 & 0.084 & $3174 - 6699$ \\
d084 & 2003jm  & 0.519 &   3.6 & -0.197 & 0.221 & $2821 - 5879$ \\
d085 & 2003jv  & 0.401 &  -0.4 & -0.228 & 0.182 & $3790 - 6795$ \\
d086 & 2003ju  & 0.205 &  -4.8 & -0.177 & 0.628 & $4929 - 7983$ \\
d086 & 2003ju  & 0.205 &  20.9 & -0.177 & 0.628 & $3070 - 8174$ \\
d087 & 2003jr  & 0.340 &  14.7 & -0.112 & 0.126 & $4044 - 7208$ \\
d089 & 2003jl  & 0.436 &  10.2 & -0.198 & 0.134 & $2973 - 6218$ \\
d093 & 2003js  & 0.363 &  -2.8 & -0.365 & 0.077 & $3117 - 6552$ \\
e142 & 2003js  & 0.363 &  15.0 & -0.365 & 0.077 & $3111 - 6555$ \\
d097 & 2003jt  & 0.436 &   4.5 & -0.317 & 0.116 & $2957 - 6218$ \\
d099 & 2003ji  & 0.211 &  18.7 & -0.110 & 0.118 & $4475 - 7976$ \\
d117 & 2003jw  & 0.309 &  -4.7 &  0.298 & 0.209 & $3254 - 6822$ \\
d149 & 2003jy  & 0.342 &  -8.1 & -0.214 & 0.170 & $3182 - 6654$ \\
e020 & 2003kk  & 0.159 &  -2.4 & -0.034 & 0.437 & $3192 - 7592$ \\
e029 & 2003kl  & 0.332 &  -1.7 &  0.219 & 0.259 & $2777 - 6606$ \\
e121 & 2003kl  & 0.332 &   0.5 &  0.219 & 0.259 & $3021 - 6906$ \\
e108 & 2003km  & 0.469 & -10.8 & -0.280 & 0.071 & $2518 - 6126$ \\
e108 & 2003km  & 0.469 & -10.1 & -0.280 & 0.071 & $2566 - 6739$ \\
e108 & 2003km  & 0.469 & -10.1 & -0.280 & 0.071 & $2952 - 6072$ \\
e132 & 2003kn  & 0.239 &  -5.7 & -0.128 & 0.739 & $3410 - 7211$ \\
e136 & 2003ko  & 0.352 &  -1.2 &  0.332 & 0.304 & $3125 - 6609$ \\
e138 & 2003kt  & 0.612 &   6.5 & -0.284 & 0.103 & $2660 - 5543$ \\
e140 & 2003kq  & 0.631 &  -1.4 & -0.187 & 0.121 & $2608 - 5478$ \\
e147 & 2003kp  & 0.645 &   1.0 & -0.174 & 0.071 & $2586 - 5432$ \\
e148 & 2003kr  & 0.429 &  -7.1 & -0.107 & 0.102 & $3010 - 6192$ \\
e149 & 2003ks  & 0.497 &  10.7 &  0.016 & 0.186 & $2873 - 5910$ \\
f011 & 2003lh  & 0.539 &   7.9 & -0.090 & 0.157 & $2067 - 6107$ \\
f041 & 2003le  & 0.561 &   4.8 & -0.301 & 0.086 & $2144 - 5952$ \\
f076 & 2003lf  & 0.410 &   3.1 &  0.175 & 0.227 & $2601 - 6259$ \\
f076 & 2003lf  & 0.410 &   3.1 &  0.175 & 0.227 & $2272 - 6462$ \\
f096 & 2003lm  & 0.412 &   3.7 &  0.171 & 0.324 & $2532 - 6575$ \\
f216 & 2003ll  & 0.599 &   6.2 & -0.104 & 0.117 & $2512 - 5809$ \\
f231 & 2003ln  & 0.619 &   5.7 & -0.247 & 0.089 & $2193 - 5752$ \\
f235 & 2003lj  & 0.422 &   3.4 &  0.165 & 0.133 & $2365 - 6535$ \\
f244 & 2003li  & 0.540 &   5.4 & -0.020 & 0.131 & $2283 - 6064$ \\
f308 & \nodata & 0.394 &   0.5 &  0.031 & 0.145 & $2482 - 6271$ \\
g005 & 2004fh  & 0.218 &   2.4 & -0.253 & 0.428 & $2921 - 7602$ \\
g050 & 2003fn  & 0.633 &  -1.2 & -0.318 & 0.128 & $3245 - 5835$ \\
g052 & 2004fm  & 0.383 &  -0.5 &  0.382 & 0.143 & $2516 - 6718$ \\
g055 & 2004fk  & 0.302 &   5.1 & -0.294 & 1.009 & $4086 - 7327$ \\
g097 & \nodata & 0.340 &  10.4 & -0.289 & 0.322 & $2973 - 6791$ \\
g120 & 2004fo  & 0.510 &  -0.9 & -0.286 & 0.186 & $2203 - 6160$ \\
g133 & \nodata & 0.421 &  19.5 & -0.351 & 0.452 & $3743 - 6713$ \\
g142 & \nodata & 0.399 &  13.8 &  0.210 & 0.523 & $2809 - 6719$ \\
g160 & 2004fs  & 0.493 &   9.8 & -0.308 & 0.194 & $3576 - 6403$ \\
g240 & \nodata & 0.687 &   9.5 & -0.163 & 0.062 & $1915 - 5498$ \\
h283 & 2004ha  & 0.502 &   1.4 &  0.090 & 0.265 & $2331 - 5872$ \\
h300 & \nodata & 0.687 &   8.5 & -0.279 & 0.076 & $1940 - 5503$ \\
h311 & 2004hc  & 0.750 &   5.9 & -0.491 & 0.065 & $2018 - 5300$ \\
h319 & 2004hd  & 0.495 &  -5.8 & -0.274 & 0.159 & $3571 - 6381$ \\
h323 & 2004he  & 0.603 &  -0.9 & -0.108 & 0.120 & $3331 - 5951$ \\
h342 & 2004hf  & 0.421 &  11.5 & -0.356 & 0.085 & $3659 - 7037$ \\
h359 & 2004hi  & 0.348 &  11.1 & -0.182 & 0.299 & $3030 - 5708$ \\
h363 & 2004hh  & 0.213 &   3.4 &  0.065 & 0.775 & $4154 - 7666$ \\
h364 & 2004hj  & 0.344 &   5.3 & -0.006 & 0.087 & $2752 - 7293$ \\
k396 & 2004hk  & 0.271 &  -5.0 &  0.843 & 0.175 & $3202 - 7633$ \\
k425 & 2004hl  & 0.274 &  10.8 & -0.021 & 0.250 & $2360 - 5767$ \\
k429 & 2004hm  & 0.181 &  -1.3 & -0.094 & 0.126 & $3234 - 8000$ \\
k430 & 2004hn  & 0.582 &   0.7 & -0.112 & 0.113 & $2261 - 5891$ \\
k441 & 2004hq  & 0.680 &   2.3 & -0.203 & 0.091 & $3142 - 5678$ \\
k448 & 2004hr  & 0.401 &   0.7 &  0.003 & 0.311 & $2872 - 5888$ \\
k485 & 2004hs  & 0.416 &   5.5 & -0.230 & 0.849 & $2895 - 6207$ \\
m026 & \nodata & 0.653 &  12.0 & -0.201 & 0.102 & $3361 - 6215$ \\
m027 & \nodata & 0.286 &   6.6 & -0.134 & 0.362 & $3383 - 6903$ \\
m027 & \nodata & 0.286 &   9.1 & -0.134 & 0.362 & $2690 - 7230$ \\
m032 & \nodata & 0.155 &  15.4 & -0.158 & 0.092 & $3688 - 8052$ \\
m034 & \nodata & 0.562 &   9.2 & -0.087 & 0.130 & $2786 - 5684$ \\
m039 & \nodata & 0.249 &  16.9 & -0.223 & 0.507 & $3482 - 7106$ \\
m039 & \nodata & 0.249 &  17.1 & -0.223 & 0.507 & $4191 - 7833$ \\
m043 & \nodata & 0.266 &  11.4 & -0.466 & 1.033 & $3436 - 7012$ \\
m057 & \nodata & 0.184 &  14.1 & -0.601 & 0.218 & $3673 - 7497$ \\
m062 & \nodata & 0.317 &   3.7 &  0.149 & 0.135 & $3721 - 7424$ \\
m138 & \nodata & 0.582 &  -6.7 & -0.306 & 0.073 & $2750 - 5612$ \\
m138 & \nodata & 0.582 &  -6.6 & -0.306 & 0.073 & $3128 - 5877$ \\
m138 & \nodata & 0.582 &  -5.9 & -0.306 & 0.073 & $2247 - 5897$ \\
m158 & \nodata & 0.463 &  10.7 & -0.334 & 0.222 & $3133 - 4812$ \\
m158 & \nodata & 0.463 &  11.8 & -0.334 & 0.222 & $3636 - 6520$ \\
m193 & \nodata & 0.341 &   6.9 & -0.124 & 0.109 & $3187 - 5249$ \\
m226 & \nodata & 0.671 &  13.9 & -0.227 & 0.113 & $3183 - 5569$ \\
n256 & \nodata & 0.631 &   6.1 & -0.339 & 0.075 & $2667 - 5441$ \\
n258 & \nodata & 0.522 &   6.0 &  0.032 & 0.110 & $2858 - 5831$ \\
n263 & \nodata & 0.368 &   2.7 &  0.054 & 0.075 & $3181 - 6488$ \\
n278 & \nodata & 0.309 &   5.6 &  0.033 & 0.156 & $3321 - 6779$ \\
n285 & \nodata & 0.528 &  13.5 & -0.122 & 0.170 & $2848 - 5808$ \\
n326 & \nodata & 0.268 &  -2.7 &  0.703 & 0.144 & $3431 - 6999$ \\
n404 & \nodata & 0.216 &  -4.3 & -0.069 & 0.694 & $3575 - 7298$ \\
p425 & \nodata & 0.453 &  11.3 &  0.183 & 0.224 & $2994 - 6108$ \\
p425 & \nodata & 0.453 &  12.1 &  0.183 & 0.224 & $3423 - 6399$ \\
p454 & \nodata & 0.695 &   2.1 & -0.276 & 0.070 & $2566 - 5236$ \\
p455 & \nodata & 0.284 & -10.6 & -0.061 & 0.229 & $3386 - 6911$ \\
p524 & \nodata & 0.508 &   9.2 & -0.229 & 0.155 & $2824 - 4668$ \\
p528 & \nodata & 0.777 &   4.1 & -0.310 & 0.041 & $2397 - 3961$ \\
p528 & \nodata & 0.777 &   5.9 & -0.310 & 0.041 & $2807 - 5228$ \\
p534 & \nodata & 0.615 &  -3.4 & -0.096 & 0.107 & $3094 - 5980$ \\
p534 & \nodata & 0.615 &  -2.3 & -0.096 & 0.107 & $3294 - 5736$ \\

\enddata

\tablenotetext{a}{Rest-frame phase relative to $B$-band maximum.}

\tablenotetext{b}{$M_{V}(t = 0) = -19.504~\textrm{mag} + 0.736\Delta +
0.182 \Delta^{2} + 5 \log_{10} (H_{0}/65)$ \citep{Jha07}.}

\end{deluxetable}


\end{document}